\documentclass[a4paper]{elsart}
\usepackage{mypaper,bm}
\date{October 24, 2004}
\newcommand{\bun}[2]{\frac{\displaystyle #1}{\displaystyle #2}}
\newcommand{\hen}[2]{\bun{\partial #1}{\partial #2}}
\newcommand{\henhen}[2]{\bun{\partial^2 #1}{\partial #2^2}}
\newcommand{\divv}{\nabla\cdot\bm{v}}
\newcommand{\pst}{\tilde{t}}
\newcommand{\dst}{\Delta\pst}
\newcommand{\reta}{r_\eta}
\newcommand{\nsmooth}{N_s}
\newcommand{\wsmooth}{W_s}
\newcommand{\nvcycle}{N_V}
\newlength{\thiswidth}
\newlength{\thisheight}
\begin{document}
%%%%%%%%%%%%%%%%%%%%%%%%%%%%%%%%% title page %%%%%%%%%%%%%%%%%%%%%%%%%%%%%%%%%%
%\input{title.tex}
\begin{frontmatter}

 \title{
 Multigrid iterative algorithm using pseudo-compressibility for
 three-dimensional mantle convection with strongly variable viscosity
 }

 \author{
 Masanori Kameyama, Akira Kageyama and Tetsuya Sato
 }

 \address{
 Earth Simulator Center,
 Japan Agency for Marine-Earth Science and Technology (JAMSTEC),
 Yokohama, 236-0001, Japan
 }

 \begin{abstract}
  A numerical algorithm for solving mantle convection problems with
  strongly variable viscosity is presented.
  Equations for conservation of mass and momentum for highly viscous and
  incompressible fluids are solved iteratively by a multigrid method in
  combination with pseudo-compressibility and local time stepping
  techniques.
  This algorithm is suitable for large-scale three-dimensional numerical
  simulations, because (i) memory storage for any additional matrix is not
  required and (ii) vectorization and parallelization are straightforward.
  The present algorithm has been incorporated into a mantle convection
  simulation program based on the finite-volume discretization in a
  three-dimensional rectangular domain.
  Benchmark comparisons with previous two- and three-dimensional
  calculations including the temperature- and/or depth-dependent
  viscosity revealed that accurate results are successfully reproduced
  even for the cases with viscosity variations of several orders of
  magnitude.
  The robustness of the numerical method against viscosity variation can
  be significantly improved by increasing the pre- and post-smoothing
  calculations during the multigrid operations, and the convergence can be
  achieved for the global viscosity variations up to $10^{10}$.
 \end{abstract}


\end{frontmatter}

%%%%%%%%%%%%%%%%%%%%%%%%%% Section 1 %%%%%%%%%%%%%%%%%%%%%%%%%%%%%%%%%%%%%%%%%%
%\input{intro.tex}
\section{Introduction}

The Earth's mantle is the spherical shell composed of silicate rocks,
and it ranges from approximately 5-50km to 2900km depth.
Although the mantle behaves like an elastic solid on short time scales,
it acts like a highly viscous fluid on long time scales.
The mantle also acts as a heat engine, and it is in a convective motion
in order to transport the heat from the hot interior to the cool
surface \cite{turcotte82,stacey92}.
The mantle convection is observed as the motion of tectonic plates at
the Earth's surface.
The motion of surface plates, in turn, drives seismicity, volcanism and
mountain building at the plate margins.
Thus, the mantle convection is the origin of the geological and
geophysical phenomena observed at the Earth's surface.
A major tool for understanding the mantle convection is numerical
analysis.
It has been playing an important role in the study of mantle convection,
since a numerical simulation of mantle convection first arose
\cite{turcotte67,mckenzie73}.

Mantle convection requires different numerical techniques from those for
ordinary fluids such as water because of its rheological properties.
%%% original %%%
%The viscosity of mantle materials is estimated as high as $10^{21}$
%{Pa s} \cite{turcotte82}.
%%% revised %%%
The viscosity of mantle materials is estimated as high as $10^{22}$
{Pa s} \cite{turcotte82,schubert01b}.
%%%
Since the mantle materials is highly viscous, both the nonlinear and
time-derivative terms of velocity can be ignored in the equation of
motion.
%%% original %%%
%This implies that the flow in the mantle is described by a steady-state
%Stokes flow balancing between the buoyancy force, pressure gradient and
%viscous resistance.
%%% revised %%%
This implies that the flow in the mantle is described by a steady-state
Stokes flow balancing among the buoyancy force, pressure gradient and
viscous resistance.
%%%
Taken together with the assumption of incompressibility, one needs to
solve elliptic differential equations for velocity and pressure at every
timestep.
In addition, the viscosity of mantle material varies by several orders
of magnitude depending on temperature, pressure, and stress
\cite{weertman70,karato93a}.
The strong variation in viscosity makes numerical techniques for
ordinary isoviscous fluids, such as the spectral method
\cite{travis90,christensen91}, unfit for the numerical modeling of
mantle convection.
In order to get deep insights into the mantle convection, it is very
important to develop efficient numerical techniques that can deal with
the steady-state flow of highly viscous and incompressible fluids with a
strongly variable viscosity.

The efficiency of numerical simulations of mantle convection strongly
relies on numerical methods used for solving elliptic differential
equations.
One of the most efficient methods is the multigrid iteration
\cite{brandt77}.
%%% original %%%
%The multigrid concept has been successfully applied to a wide range of
%problems, including calculations of incompressible fluid flow
%\cite{brandt80a,wesseling-book}.
%%% revised %%%
The multigrid concept has been successfully applied to a wide range of
problems, including calculations of incompressible fluid flow
\cite{brandt80a,wesseling-book,trottenberg01}.
%%%
During the last two decades, various numerical models of mantle
convection have been developed where the multigrid method is utilized.
There are two strategies to apply the multigrid method to this problem,
depending on how the steady-state Stokes equations are solved.

The first strategy solves the Stokes equations by splitting into the
separate equations for velocities and pressure.
The discretized equations for velocity components (or their proxy) are
solved by the multigrid method, while the pressure is eliminated or
solved separately.
Parmentier et al.~\cite{parmentier94} developed convection models of
isoviscous fluid in three-dimensional Cartesian geometry.
By using a streamfunction formulation, the Stokes equations are reduced
to a pair of Poisson equations which are solved by multigrid iterations.
Baumgardner and his colleagues developed convection models for fluids
with constant \cite{baumgardner85} and moderately variable viscosity
\cite{bunge96a,bunge98} in a three-dimensional spherical geometry.
They solved the elliptic equations for velocity components using a
multigrid method, while the pressure fields are prescribed by the
equation of state.
Moresi and his colleagues \cite{moresi95a,moresi96a} developed a model
for convection problems with strongly variable viscosity in two- and
three-dimensional Cartesian geometry.
The Stokes equations are solved separately for velocity and pressure by
so-called Uzawa iterative scheme.
The iteration for velocity is carried out by a multigrid method, while a
conjugate gradient scheme is used for pressure iteration.

The second strategy, on the other hand, solves the Stokes equations for
velocity and pressure as a whole by the multigrid technique.
The key issue of this strategy is a choice of the smoothing algorithm
which reduces the errors of solution on a particular grid.
Several methods for solving incompressible fluid flows have been
utilized as a smoothing algorithm.
Trompert and Hansen \cite{trompert96,trompert98a} and Albers
\cite{albers00} developed numerical methods for convection problems with
variable viscosity in three-dimensional Cartesian geometry.
The Stokes equations are solved by a multigrid method where the SIMPLER
algorithm \cite{patankar80} is employed as a smoothing operation.
Auth and Harder \cite{auth99} used the symmetric coupled Gauss-Seidel
(SCGS) method \cite{vanka-scgs} as a smoothing operation in the
multigrid method for solving the Stokes equation.
Tackley \cite{tackley-thesis,tackley96b} employed a similar smoothing
method where the velocity and pressure fields are updated in a somewhat
coupled manner, and his method has been successfully applied to a
variety of convection problems \cite{tackley96b,ratcliff97}, including
calculations with extremely variable viscosity
\cite{tackley98,tackley00b,tackley00c}.

In this paper, we present a solution algorithm for the three-dimensional
mantle convection, following the second strategy where the Stokes
equations are solved as a whole by the multigrid method.
In Section \ref{method-section} we introduce our solution algorithm
based on the pseudo-compressibility method \cite{chorin67}.
A notable feature of our algorithm is its simplicity and intuitiveness.
This feature makes the algorithm easily fit to a smoothing operation
of the multigrid iterations.
In addition, taken together with the local time-stepping method, a
strong variation in viscosity can be handled without a severe increase
in computational costs.
In Section \ref{result-section}, we develop a numerical model of mantle
convection that can deal with a strongly variable viscosity based on the
algorithm presented in Section \ref{method-section} together with the
multigrid technique.
In order to demonstrate the validity and efficiency of our algorithm, we
carry out several calculations for the mantle convection with a strongly
variable viscosity.
In this paper, an application of our algorithm is the thermal convection
in a three-dimensional rectangular domain.
It can be easily applied to the numerical models for more realistic
problems of mantle convection, such as spherical shell geometry and
thermo-chemical convection and so on.

%\input{method.tex}
%%%%%%%%%%%%%%%%%%%%%%%%%% Section 2 %%%%%%%%%%%%%%%%%%%%%%%%%%%%%%%%%%%%%%%%%%
\section{Iterative Algorithm for mantle convection}
\label{method-section}

%%% original %%%
%We consider a thermal convection of a highly viscous and incompressible
%fluid with a strongly variable Newtonian viscosity in a
%three-dimensional Cartesian geometry ($x_1$=$x$, $x_2$=$y$, $x_3$=$z$).
%%% revised %%%
We consider thermal convection of a highly viscous and incompressible
fluid with a strongly variable Newtonian viscosity in a
three-dimensional Cartesian geometry ($x_1$=$x$, $x_2$=$y$, $x_3$=$z$).
%%%
The nondimensional forms of the fundamental equations are
\cite[for example]{schmeling81,ogawa91};
\begin{eqnarray}
 -\hen{}{x_j}\left[\eta\left(\hen{v_j}{x_i}+\hen{v_i}{x_j}\right)\right] + \hen{p}{x_i} & = & Ra\,T\delta_{i3} \quad(i,j=1,2,3),
  \label{mom_eq}
  \\
 \hen{v_k}{x_k} & = & 0 \quad(k=1,2,3),
  \label{cont_eq}
  \\
 \hen{T}{t}+v_i\hen{T}{x_i} & = & \henhen{T}{x_j}+q,
  \label{heat_eq}
\end{eqnarray}
where $v_i$ ($i=1,2,3$) are fluid velocities in $i$-th direction, $p$
pressure, $\eta$ viscosity, $T$ temperature, $Ra$ the Rayleigh number,
and $q$ is the internal heating rate.
We assumed that $x_3$-axis is the vertical axis pointing upward.
%%% original %%%
%Boussinesq approximation is employed in the energy equation
%(\ref{heat_eq}) and, hence, the effects of adiabatic and viscous
%heatings are ignored.
%%% revised %%%
Boussinesq approximation is employed in the energy equation
(\ref{heat_eq}) and, hence, the effects of adiabatic and viscous
heating are ignored.
%%%

\subsection{Overview of the iterative algorithm}

In the following, we consider the solution algorithm for the
conservation equations of momentum (\ref{mom_eq}) and mass
(\ref{cont_eq}).
In order to simplify the solution algorithm, the energy equation
(\ref{heat_eq}) is solved separately from Eqs.~(\ref{mom_eq}) and
(\ref{cont_eq}): the temperature $T$ at new time is calculated using the
velocity at old time in the advection term.
This algorithm solves Eqs.~(\ref{mom_eq}) and (\ref{cont_eq}) for the
velocity and pressure at new time using the new temperature.
In addition, the viscosity $\eta$ at new time is already known by some
means and kept unchanged while the velocity and pressure are computed.
%%% original %%%
%Because of these assumptions, Eqs.~(\ref{mom_eq}) and (\ref{cont_eq})
%are taken to be linear with respect to $v_1$ to $v_3$ and $p$.
%%% revised %%%
Because of these assumptions, Eqs.~(\ref{mom_eq}) and (\ref{cont_eq})
are taken to be linear with respect to velocity and pressure.
%%%

The discretized equations of (\ref{mom_eq}) and (\ref{cont_eq}) can be
written in a matrix form of
\begin{equation}
 \bm{A}\bm{x}=\bm{b},
  \label{ax=b}
\end{equation}
where
\begin{eqnarray}
 \bm{A}\equiv
  \left[
   \begin{array}{cccc}
    \bm{C}_{11} & \bm{C}_{12} & \bm{C}_{13} & \bm{G}_1 \\
    \bm{C}_{21} & \bm{C}_{22} & \bm{C}_{23} & \bm{G}_2 \\
    \bm{C}_{31} & \bm{C}_{32} & \bm{C}_{33} & \bm{G}_3 \\
    \bm{D}_{1} & \bm{D}_{2} & \bm{D}_{3} & \bm{0} \\
   \end{array}
 \right]
  ,\quad
 \bm{x}\equiv
  \left[
   \begin{array}{c}
    \bm{v}_1 \\ \bm{v}_2 \\ \bm{v}_3 \\ \bm{p} \\
   \end{array}
 \right]
  ,\quad
 \bm{b}\equiv
  \left[
   \begin{array}{c}
    \bm{0} \\ \bm{0} \\ Ra\,\bm{\theta} \\ \bm{0} \\
   \end{array}
 \right].
  \label{def-abx}
\end{eqnarray}
In (\ref{def-abx}), $\bm{v}_{i}$ $(i=1,2,3)$ are the vectors of unknown
fluid velocities in $i$-direction, $\bm{p}$ the vector of unknown
pressures, $\bm{\theta}$ the vector of known temperatures, $\bm{C}_{ij}$
$(i,j=1,2,3)$ the matrices representing the spatial discretization of
velocities for calculating viscous stress, $\bm{G}_{i}$ $(i=1,2,3)$ the
matrices representing the discretization of the pressure gradient,
$\bm{D}_{i}$ $(i=1,2,3)$ the matrices representing the discretization of
the divergence of velocity, and $\bm{0}$ is the zero matrix or zero
vector.
One of the major difficulties in solving (\ref{ax=b}) is that a zero
block appears in the main diagonal in $\bm{A}$.
This implies that basic iterative methods for solving linear
simultaneous equations (such as Jacobi and Gauss-Seidel methods) which
make use of the inverse of the main diagonal fail to solve
(\ref{ax=b}).

The algorithm for solving (\ref{ax=b}) that we propose in this paper is
another kind of iterative method:
Let $\bm{x}^n$ be an approximate at $n$-th iteration
($n=0,1,2,\cdots$).
%%% original %%%
%We then calculate the approximate at $(n+1)$-th iteration by,
%%% revised %%%
We then calculate the approximate at the iteration $(n+1)$ by,
%%%
\begin{equation}
 \bm{x}^{n+1} = \bm{x}^{n}+\bm{T}(\bm{b}-\bm{A}\bm{x}^{n}).
  \label{ax=b-iter}
\end{equation}
Here a regular matrix $\bm{T}$ is introduced so as to control the
convergence rate of (\ref{ax=b-iter}), and is chosen to be a diagonal
matrix whose nonzero elements are positive.
This iterative procedure is used together with the multigrid method, and
is repeated until $\bm{x}^n$ converges to a steady solution.
It is obvious that the steady solution of (\ref{ax=b-iter}) satisfies
(\ref{ax=b}), and that it does not depend on $\bm{T}$.
In addition, the presence of zero block in the main diagonal of $\bm{A}$
does not spoil the convergence of this iterative procedure.

This iterative algorithm is suitable for large-scale numerical
simulations on massively parallel computers and, in particular,
massively vector-parallel supercomputers because (i) the procedure
(\ref{ax=b-iter}) consists of multiplication of matrices and vectors and
summation of vectors and, hence, can be easily vectorized and
parallelized, and (ii) large memory storage for the matrix $\bm{T}$ is
not required, since $\bm{T}$ is a diagonal matrix.

Another benefit of the iterative procedure (\ref{ax=b-iter}) is that it
is easily implemented into the multigrid method.
This is because the procedure (\ref{ax=b-iter}) is quite similar to that
representing the basic iterative methods which are commonly used as a
smoothing operator of the multigrid method.
Suppose an iterative procedure for (\ref{ax=b}) by Jacobi method;
\begin{equation}
 \bm{x}^{n+1}=\bm{x}^n+\bm{D}^{-1}(\bm{b}-\bm{A}\bm{x}^n)
  \quad (n=0,1,2\cdots),
  \label{jacobi-method}
\end{equation}
where a matrix $\bm{D}$ is the main diagonal of $\bm{A}$.
The two procedures (\ref{ax=b-iter}) and (\ref{jacobi-method}) are the
same except that the diagonal matrix $\bm{T}$ is used in
(\ref{ax=b-iter}) in place of $\bm{D}^{-1}$ in (\ref{jacobi-method}).
Therefore, the iterative procedure (\ref{ax=b-iter}) can be used as a
smoothing operator of the multigrid method in a similar manner to Jacobi
method.

In the following subsections, we will introduce the ideas which led us
to the iterative procedure (\ref{ax=b-iter}).
We will also discuss an appropriate choice of the matrix $\bm{T}$, which
is a key parameter in this procedure.

%%%%%%%%%%%%%%%%%%%%%%%%%% Section 2.1 %%%%%%%%%%%%%%%%%%%%%%%%%%%%%%%%%%%%%%%%
%\input{pcm.tex}
\subsection{Ingredient 1: pseudo-compressibility method for
  highly-viscous, incompressible fluids}

When the matrix $\bm{T}$ is diagonal, Eq.~(\ref{ax=b-iter}) can be
explicitly written as ($i,j,k=1,2,3$),
\begin{eqnarray}
 v_i^{n+1} & = & v_i^{n} + \tau_{v_i}
  \left\{
   -\hen{p^n}{x_i} + \hen{}{x_j}\left[\eta\left(\hen{v_j^n}{x_i}+\hen{v_i^n}{x_j}\right)\right] + Ra\,T\delta_{i3}
\right\},
\label{mom_iter}
\\
 p^{n+1} & = & p^{n} - \tau_p\hen{v_k^n}{x_k},
  \label{cont_iter}
\end{eqnarray}
where $\tau_{v_i}$ ($i=1,2,3$) and $\tau_p$ are the diagonal elements
appearing in $\bm{T}$.

In order to discuss the physical meaning of these equations more
clearly, we introduce the following ``auxiliary'' set of equations;
\begin{eqnarray}
 M_{v_i}\hen{v_i}{\pst} & = & -\hen{p}{x_i} +\hen{}{x_j}\left[\eta\left(\hen{v_j}{x_i}+\hen{v_i}{x_j}\right)\right] + Ra\,T\delta_{i3},
  \label{mom_evol}
  \\
 K\hen{p}{\pst}  & = & -\hen{v_k}{x_k}.
  \label{cont_evol}
\end{eqnarray}
Here $\pst$ is analogous to time, $M_{v_i}$ ($i=1,2,3$) and $K$ are
positive constants analogous to density and compressibility,
respectively.
These equations come from the modification of the ``auxiliary'' set of
equations used in the pseudo-compressibility method \cite{chorin67},
which is used to solve steady-state flow of incompressible fluid with
high Reynolds number.
The difference between Eqs.~(\ref{mom_evol}) and (\ref{cont_evol}) and
those employed for high Reynolds-number flow is that the nonlinear term
of velocity has been eliminated in (\ref{mom_evol}), because the
viscosity of mantle materials is significantly high and, in other words,
the Reynolds number is significantly small.

By discretizing (\ref{mom_evol}) and (\ref{cont_evol}) in the direction
of $\pst$ by a first-order explicit scheme, we get
\begin{eqnarray}
 M_{v_i}\bun{v_i^{n+1}-v_i^n}{\dst} & = & -\hen{p^n}{x_i} + \hen{}{x_j}\left[\eta\left(\hen{v_j^n}{x_i}+\hen{v_i^n}{x_j}\right)\right] +RaT\delta_{i3},
  \label{mom_pcm}
  \\
 K\bun{p^{n+1}-p^n}{\dst}  & = & -\hen{v_k^n}{x_k}.
  \label{cont_pcm}
\end{eqnarray}
We notice that Eqs.~(\ref{mom_pcm}) and (\ref{cont_pcm}) and
Eqs.~(\ref{mom_iter}) and (\ref{cont_iter}) are identical if we choose
$\tau_{v_i}$ and $\tau_p$ as,
\begin{equation}
 \tau_{v_i}=\dst/M_{v_i}, \quad \tau_p=\dst/K.
  \label{pseudotimestep}
\end{equation}
In other words, $\tau_{v_i}$ and $\tau_p$ are ``effective'' timesteps
for the evolution of $v_i$ and $p$, respectively.

The iterative procedure given by (\ref{mom_pcm}) and (\ref{cont_pcm})
converges to an incompressible flow field regardless of the choice of
$M_{v_i}$ and $K$.
To show this more clearly, we here assume that $M_{v_i}$, $K$ and $\eta$
are constant.
From Eqs.~(\ref{mom_evol}) and (\ref{cont_evol}) we obtain the
pseudo-temporal evolution of $\divv$ as,
\begin{equation}
 \henhen{}{\pst}(\divv) = \bun{1}{KM}\nabla^2(\divv)+\bun{2\eta}{M}\hen{}{\pst}\left[\nabla^2\left(\divv\right)\right].
  \label{divv_evol}
\end{equation}
%%% original %%%
%(Here we denoted $M_{v_i}\equiv M$.)
%%% revised %%%
(Here we denote $M_{v_i}\equiv M$.)
%%%
The first term of the right-hand side represents the effect of
propagation of ``pseudo-sound wave'' whose velocity is $1/\sqrt{KM}$,
and the second term represents the effect of diffusion due to
viscosity.
This equation indicates that the pseudo-temporal evolution of $\divv$ is
characterized by a decaying oscillation.
Therefore, $\divv$ approaches to an asymptotic value ($=0$) as $\pst$
increases to infinity.
Consequently, the procedure of (\ref{mom_pcm}) and (\ref{cont_pcm})
leads the steady velocity field with $\divv=0$, as long as the numerical
integration scheme is stable.

We also note that the iterative procedure (\ref{mom_iter}) and
(\ref{cont_iter}) should be used in combination with multigrid method
\cite{brandt77}.
Recall that we are trying to find a steady-state solution of evolution
equations (\ref{mom_evol}) and (\ref{cont_evol}) through a repetition of
temporal integration, which inevitably requires a large number of
iterations.
Moreover, the elliptic nature of (\ref{mom_evol}) for velocities $v_i$
results in a slow reduction in their errors, particularly in their
smooth components with large spatial wavelengths.
%%% original %%%
%By incorporating into multigrid operations, we are able to obtain a
%sufficiently fast convergence.
%%% revised %%%
By incorporating into the multigrid procedure, we are able to obtain a
sufficiently fast convergence.
%%%
In fact, the multigrid technique is adopted in most of recent numerical
analysis using pseudo-compressibility method \cite[for
example]{sotiropoulos97,tai03}.

%%%%%%%%%%%%%%%%%%%%%%%%%% Section 2.2 %%%%%%%%%%%%%%%%%%%%%%%%%%%%%%%%%%%%%%%%
%\input{lts.tex}
\subsection{Ingredient 2: local time stepping method for strongly
  variable viscosity}\label{lts}

In most of previous numerical analysis using pseudo-compressibility
method \cite[for example]{sotiropoulos97,tai03,rogers87,shen96}, the
viscosity of fluids has been assumed to be constant.
In this subsection, we apply this method for the cases with a strong
variation in viscosity.

The convergence rate of the original pseudo-compressibility method
deteriorates in proportion to the viscosity variation in the entire
domain, when the spatial variation in viscosity is introduced.
This is because the viscosity $\eta$ is a diffusion coefficient for
$v_i$ in Eq.~(\ref{mom_evol}).
The spatial variation in $\eta$ results in the spatial variation in the
local convergence rate of $v_i$.
%%% original %%%
%The values of $v_i$ in the region with smaller $\eta$ reach to an
%appropriate solution only very slowly, whereas the values in the region
%with larger $\eta$ quickly do.
%%% revised %%%
The values of $v_i$ in the region with smaller $\eta$ reach to an
appropriate solution only very slowly, whereas the values in the region
with larger $\eta$ converge quickly.
%%%
This suggests that the convergence should be accelerated particularly in
the region with small $\eta$.

Here we try to accelerate the convergence for the case with a spatial
variation in $\eta$ by controlling the stepping of the pseudo-time $\pst$
in accordance with $\eta$.
This idea is known as the local time stepping method, which is used to
obtain steady-state solution of evolutionary equations.
This method allows updating each variable using a timestep which is
based on the local numerical stability criterion.
We thus spatially vary the effective timesteps $\tau_{v_i}$ and $\tau_p$
in (\ref{mom_iter}) and (\ref{cont_iter}) in accordance with $\eta$.

The spatial variation of $\tau_{v_i}$ and $\tau_p$ can be estimated from
the local numerical stability criterion of ($\ref{divv_evol}$).
In the following derivation, we assume for simplicity that
Eq.~(\ref{divv_evol}) holds even in the case of variable viscosity.
%%% original %%%
%From the criterion of the diffusion of $\divv$ (the second term in the
%right-hand side of (\ref{divv_evol})) we make a nondimensional parameter
%$\alpha$ defined by
%%% revised %%%
From the criterion of the diffusion of $\divv$ (the second term in the
right-hand side of (\ref{divv_evol})) we require that the nondimensional
parameter $\alpha$ defined by
%%%
\begin{equation}
 \alpha\equiv\bun{2\eta/M}{\Delta^2/\dst}
  \label{diffusion-criterion}
\end{equation}
%%% original %%%
%be sufficiently small.
%Here $\Delta$ is a mesh size.
%%% revised %%%
is sufficiently small.
Here $\Delta$ is the mesh size.
%%%
%%% original %%%
%Similarly, from the criterion of the pseudo-sound propagation we make a
%nondimensional parameter $\beta$ defined by
%%% revised %%%
Similarly, from the criterion of the pseudo-sound propagation we require
that the nondimensional parameter $\beta$ defined by
%%%
\begin{equation}
 \beta\equiv\bun{1/\sqrt{KM}}{\Delta/\dst}
  \label{wave-criterion}
\end{equation}
%%% original %%%
%be sufficiently small.
%%% revised %%%
is sufficiently small.
%%%
By substituting (\ref{diffusion-criterion}) and (\ref{wave-criterion})
into (\ref{pseudotimestep}) we get
\begin{equation}
 \tau_{v_i} = \bun{\dst}{M}=\bun{\alpha}{2}\bun{\Delta^2}{\eta},
  \quad
  \tau_p = \bun{\dst}{K}=\bun{2\beta^2}{\alpha}\eta.
  \label{ltsdef}
\end{equation}
Here we assumed $M_{v_i}=M$.

When $\tau_{v_i}$ and $\tau_p$ are defined by (\ref{ltsdef}), the
spatial variations of the diffusion of velocity components and the
pseudo-sound propagation become more modest than that of viscosity.
Equation (\ref{ltsdef}) implies that $M$ ($\equiv M_{v_i}$ in
(\ref{mom_evol})) and $K$ are taken to be proportional to $\eta$ and
$\eta^{-1}$, respectively.
By assuming that $\eta/M$ and $K\eta$ are constant in (\ref{mom_evol})
and (\ref{cont_evol}), we obtain the pseudo-temporal evolution of
$\divv$ in the presence of spatial variation in $\eta$ as,
\begin{eqnarray}
 \henhen{}{\pst}(\divv) & = &
  \bun{1}{KM}\nabla^2(\divv)+\bun{2\eta}{M}\hen{}{\pst}\left[\nabla^2(\divv)\right]
  \nonumber\\
 & + &
  \bun{1}{KM}\hen{}{x_i}\left[\hen{(\ln\eta)}{x_i}(\divv)\right]
  \nonumber\\
 & + &
  \bun{\eta}{M}\hen{}{x_i}\left[\hen{(\ln\eta)}{x_j}\hen{}{\pst}\left(\hen{v_j}{x_i}+\hen{v_i}{x_j}\right)\right].
  \label{divv_evol_variable_eta}
\end{eqnarray}
The first and second terms in the right-hand side of
(\ref{divv_evol_variable_eta}) are equivalent with those in the
right-hand side of (\ref{divv_evol}).
Note that the effects of spatial variation in $\eta$ appear in the third
and fourth terms in the form of $\ln\eta$.
Thus we expect that $\divv$ reduces to zero in the entire computational
domain at a rate comparable to that for the case with constant $\eta$.

Another constraint for $\tau_{v_i}$ and $\tau_p$ can be obtained
directly from (\ref{ax=b-iter}).
Let $\bm{e}^n\equiv\bm{x}^n-\bm{A}^{-1}\bm{b}$ and
$\bm{r}^n\equiv\bm{b}-\bm{A}\bm{x}^n$ to be the error and the residual
at $n$-th iteration, respectively.
From (\ref{ax=b-iter}), we get the evolution equations for $\bm{e}^n$
and $\bm{r}^n$ as,
\begin{equation}
 \bm{e}^{n+1} = (\bm{I}-\bm{T}\bm{A})\bm{e}^{n},
  \quad
  \bm{r}^{n+1} = (\bm{I}-\bm{A}\bm{T})\bm{r}^{n},
  \label{evolve-error}
\end{equation}
where $\bm{I}$ is the identity matrix.
These equations indicate that the convergence rate of (\ref{ax=b-iter})
becomes optimal when $\bm{T}=\bm{T}_{\mathrm{opt}}\equiv\bm{A}^{-1}$.
(In this sense, the matrix $\bm{T}$ can be regarded as a
``preconditioner'' for Eq.~(\ref{ax=b}).)
In the present algorithm using local time stepping, we approximate
$\bm{T}\simeq\bm{T}_{\mathrm{opt}}$ by a diagonal matrix.
In addition, Equation (\ref{evolve-error}) implies that the necessary
condition for the convergence is that the spectral radius of the matrix
$\bm{I}-\bm{T}\bm{A}$ (or, identically, $\bm{I}-\bm{A}\bm{T}$) is
smaller than one.
Indeed, this condition is identical to the Courant-Friedrichs-Lewy (CFL)
condition of Eqs.~(\ref{mom_pcm}) and (\ref{cont_pcm}), i.e.,
$\alpha<\alpha_c$ and $\beta<\beta_c$ where $\alpha_c$ and $\beta_c$ are
threshold values.

%%%%%%%%%%%%%%%%%%%%%%%% Section 3 %%%%%%%%%%%%%%%%%%%%%%%%%%%%%%%%%%%%%%%%%%%%
%\input{results.tex}
\section{Results}\label{result-section}

The algorithm described in the previous section has been incorporated
into a mantle convection simulation program in a three-dimensional
rectangular domain.
We carried out several calculations for thermal convection with strongly
variable viscosity, in order to demonstrate the validity and efficiency
of the iterative algorithm.

%%%%%%%%%%%%%%%%%%%%%%%%%% Section 3.1 %%%%%%%%%%%%%%%%%%%%%%%%%%%%%%%%%%%%%%%%
%\input{codes.tex}
\subsection{Details of numerical method}

The basic equations are discretized by the finite volume method.
A staggered grid is used; temperature and pressure are located at the
center of the grid cells, while the velocity components are at the
center of the cell faces normal to the direction of the velocity
components.
A uniform mesh is employed.
The basic equations are nondimensionalized with a length scale of $h$
(height of the box), time scale of $h^2/\kappa$ (where $\kappa$ is
thermal diffusivity), temperature scale of $\Delta T$ (temperature drop
across the box), and pressure scale of $\eta_o\kappa/h^2$ (where
$\eta_o$ is reference viscosity).
In the following subsections, we present two kinds of calculations.
First is to obtain the instantaneous flow fields for prescribed
distributions of buoyancy and viscosity, and second is to obtain the
temporal developments of thermal convection.

When the purpose is to obtain the flow fields, we only solve the
conservation equations for momentum (\ref{mom_eq}) and mass
(\ref{cont_eq}).
These equations are solved by the multigrid method based on
correction-storage algorithm, because of the linear nature of
(\ref{ax=b}).
The smoothing operation at each grid level is carried out by
(\ref{ax=b-iter}).
The diagonal elements of $\bm{T}$ for unknown velocities ($\tau_{v_i}$)
were chosen to be 0.5 times the inverse of the corresponding diagonal
elements of $\bm{A}$, while those for unknown pressures ($\tau_p$) are
to be 0.25 times the viscosity at the corresponding cell centers.
The values of $\tau_{v_i}$ and $\tau_p$ defined thus are approximately
equal to those with $\alpha=\beta=\bun{1}{8}$ in (\ref{ltsdef}).
%%% revised %%%
(We observed that the iteration by (\ref{ax=b-iter}) diverged when
$\tau_{v_i}$ or $\tau_p$ is larger than the above values.)
%%%
A linear interpolation is used in both fine-to-coarse (restriction) and
coarse-to-fine (prolongation) operations.
The values of viscosity at cell centers on coarser grids is calculated
by a linear interpolation from those on the finer grids by one grid
level.
On the other hand, the values of viscosity at the midpoints of cell
edges on each grid level, which are necessary to calculate viscous shear
stress, are calculated by an interpolation proposed by Ogawa et
al.~\cite{ogawa91}
from the values at cell centers on the same grid level.
Since the aim of this paper is to demonstrate the efficiency and
robustness of the smoothing algorithm (\ref{ax=b-iter}), only the
V-cycle is used in the present calculations, although more complicated
multigrid iteration cycles (such as W-, F-cycles) are expected to be
more robust \cite[for example]{auth99,albers00}.

When the purpose is to obtain a time-dependent or steady-state
convection, we solve the energy equation (\ref{heat_eq}) in addition to
Eqs.~(\ref{mom_eq}) and (\ref{cont_eq}).
In the calculations presented in this paper, the internal heating rate
$q$ is taken to be zero.
The energy equation is discretized by a first-order Euler method in
time.
An upwind scheme, called power-law scheme \cite{patankar80}, is used to
evaluate the contributions of heat transport by advection and
conduction.
The discretized equation for $T$ is solved by a fully implicit scheme
when we seek for a steady-state convection, or by a fully explicit
scheme when we seek for a time-dependent convection.

%%%%%%%%%%%%%%%%%%%%%%%%%%%% Section 3.2 %%%%%%%%%%%%%%%%%%%%%%%%%%%%%%%%%%%%%%
%\input{benchmarks.tex}
\subsection{Benchmark comparison}

In order to test the validity of our numerical code, we compare our
results with those of earlier studies
\cite{blankenbach89,busse93,ogawa91}.
The temperature $T$ is fixed at 0 and 1 at the top ($z=1$) and bottom
($z=0$) boundaries, respectively, while the vertical side walls are
adiabatic.
In most cases, the viscosity $\eta$ is assumed to be dependent on
temperature $T$ and depth ($1-z$) as,
\begin{equation}
 \eta = \eta_t\exp[-E_T T + E_p(1-z)],
\end{equation}
where $\eta_t$ is the viscosity at the top surface ($z=1$, $T=0$), and
$E_T$ and $E_p$ are constants describing the temperature- and
depth-dependence of viscosity, respectively.

%%%%%%%%%%%%%%%%%%%%%%%%%%% Section 3.2.1 %%%%%%%%%%%%%%%%%%%%%%%%%%%%%%%%%%%%%
\subsubsection{Benchmark for two-dimensional convection}

 \begin{table}
  \caption{Benchmark comparison with Blankenbach et
  al.~\cite{blankenbach89}. See the text as for the meaning of the
  symbols.}
  \begin{center}
   \begin{tabular}{cccccc}
    \hline
    mesh & 16$\times$16 & 32$\times$32 & 64$\times$64 & 128$\times$128 &
    benchmark standard \\\hline
    \multicolumn{6}{l}{(a) Case 1a (constant viscosity, $Ra=10^4$,
    $\lambda=1$)}\\
    %mesh & 16$\times$16 & 32$\times$32 & 64$\times$64 & 128$\times$128 &
    %benchmark standard \\\hline
    $Nu$ & 4.726885 & 4.840006 & 4.872745 & 4.881429 &
    4.884409$\pm$0.000010 \\
    $V_{\mathrm{rms}}$ & 41.755217 & 42.560118 & 42.785441 & 42.844667 &
    42.864947$\pm$0.000020 \\
    $q_1$ ($=q_3$) & 7.695426 & 7.966605 & 8.036044 & 8.053536 &
    8.059384$\pm$0.000003\\
    $q_2$ ($=q_4$) & 0.887088 & 0.664927 & 0.608154 & 0.593695 &
    0.588810$\pm$0.000003 \\
    \hline
    %\multicolumn{6}{c}{}\\
    \multicolumn{6}{l}{(b) Case 1b (constant viscosity, $Ra=10^5$,
    $\lambda=1$)}\\
    %mesh & 16$\times$16 & 32$\times$32 & 64$\times$64 & 128$\times$128 &
    %benchmark standard \\\hline
    $Nu$ & 9.235550 & 10.082608 & 10.394786 & 10.495490 &
    10.534095$\pm$0.000010 \\
    $V_{\mathrm{rms}}$ & 181.18346 & 189.19219 & 191.98526 & 192.87381 &
    193.21454$\pm$0.00010 \\
    $q_1$ ($=q_3$) & 14.424512 & 17.764322 & 18.731196 & 18.990358 &
    19.079440$\pm$0.000040 \\
    $q_2$ ($=q_4$) & 3.441283 & 1.425553 & 0.898253 & 0.767465 &
    0.722751$\pm$0.000020 \\
    \hline
    %\multicolumn{6}{c}{}\\
    \multicolumn{6}{l}{(c) Case 1c (constant viscosity, $Ra=10^6$,
    $\lambda=1$)}\\
    %mesh & 16$\times$16 & 32$\times$32 & 64$\times$64 & 128$\times$128 &
    %benchmark standard \\\hline
    $Nu$ & 13.505592 & 18.963003 & 20.884342 & 21.604074 &
    21.972465$\pm$0.000020 \\
    $V_{\mathrm{rms}}$ & 675.69469 & 780.80943 & 814.68028 & 827.43096 &
    833.98977$\pm$0.00020 \\
    $q_1$ ($=q_3$) & 16.63581 & 31.21463 & 41.44228 & 44.72368 &
    45.96425$\pm$0.00030 \\
    $q_2$ ($=q_4$) & 9.836338 & 6.941879 & 2.477731 & 1.255997 &
    0.877170$\pm$0.000010 \\
    \hline
    %\multicolumn{6}{c}{}\\
    \multicolumn{6}{l}{(d) Case 2a ($Ra_t=10^4$, $E_T=\ln(10^3)$, $E_p=0$,
    $\lambda=1$)}\\
    %mesh & 16$\times$16 & 32$\times$32 & 64$\times$64 & 128$\times$128 &
    %benchmark standard \\\hline
    $Nu$ & 10.3634 & 10.4780 & 10.1666 & 10.0862 &
    10.0660$\pm$0.00020 \\
    $V_{\mathrm{rms}}$ & 372.2759 & 457.4738 & 475.5292 & 478.9647 &
    480.4334$\pm$0.1000 \\
    $q_1$ & 17.42173 & 20.41169 & 18.33101 & 17.72910 &
    17.53136$\pm$0.00400 \\
    $q_2$ & 2.25932 & 1.16135 & 1.03954 & 1.01498 &
    1.00851$\pm$0.00020 \\
    $q_3$ & 14.1505 & 20.2221 & 25.3614 & 26.6611 &
    26.8085$\pm$0.0100 \\
    $q_4$ & 7.036967 & 4.248769 & 1.557456 & 0.743177 &
    0.497380$\pm$0.000100 \\
    \hline
    mesh & 40$\times$16 & 80$\times$32 & 160$\times$64 & 320$\times$128 &
    benchmark standard \\\hline
    %\multicolumn{6}{c}{}\\
    \multicolumn{6}{l}{(e) Case 2b ($Ra_t=10^4$, $E_T=\ln(16384)$,
    $E_p=\ln(64)$, $\lambda=2.5$)}\\
    %mesh & 40$\times$16 & 80$\times$32 & 160$\times$64 & 320$\times$128 &
    %benchmark standard \\\hline
    $Nu$ & 7.4020 & 7.2761 & 7.0450 & 6.9605 &
    6.9299$\pm$0.0005 \\
    $V_{\mathrm{rms}}$ & 201.478 & 171.085 & 175.343 & 172.806 &
    171.755$\pm$0.020 \\
    $q_1$ & 18.4789 & 30.6337 & 20.9644 & 19.0478 &
    18.4842$\pm$0.0100 \\
    $q_2$ & 0.37860 & 0.17729 & 0.18048 & 0.17873 &
    0.17742$\pm$0.00003 \\
    $q_3$ & 14.5474 & 14.7261 & 14.7092 & 14.3514 &
    14.1682$\pm$0.0050 \\
    $q_4$ & 3.51882 & 1.28403 & 0.79348 & 0.66355 &
    0.61770$\pm$0.00005 \\
    \hline
   \end{tabular}
  \end{center}
  \label{blank-table}
 \end{table}

First we present the benchmark comparison with Blankenbach et
al.~\cite{blankenbach89} for the cases of steady-state convection in
two-dimensional rectangular domains of aspect ratio (width/depth)
$\lambda$.
We carried out calculations of five cases listed in Table
\ref{blank-table}.
Cases 1a to 1c are the convection of isoviscous fluid in a square box
($\lambda=1$), Case 2a is the convection of fluids with
temperature-dependent viscosity in a square box, and Case 2b is the
convection with temperature- and depth-dependent viscosity in a box of
$\lambda=2.5$.
The impermeable and shear-stress-free conditions are adopted along the
all boundaries.
The initial conditions are chosen to make single convective cell with an
ascending flow along the left-side wall ($x=0$).
We carried out these calculations by varying the number of mesh
divisions in $x$- and $z$-directions, while the mesh division in
$y$-direction was kept to 1, in order to obtain two-dimensional flow
patterns in the $x$-$z$ plane.

We summarized in Table \ref{blank-table} the parameter values and mesh
divisions employed in the present calculations, and the obtained values
of (i) Nusselt number $Nu$, (ii) root-mean-square velocity
$V_{\mathrm{rms}}$, and (iii) vertical temperature gradient $q_1$ to
$q_4$ at $(x,z)=(0,1)$, $(\lambda,1)$, $(\lambda,0)$, and $(0,0)$,
respectively.
We also show in Table \ref{blank-table} the results from the benchmark
standard \cite{blankenbach89}.
Table \ref{blank-table} shows that the agreement between the results of
our calculations and benchmark standards is satisfactory for all the
values.
We conclude that the present numerical code accurately handles
two-dimensional convection of fluids with both constant and variable
viscosity.

%%%%%%%%%%%%%%%%%%%%%%%%%% Section 3.2.2 %%%%%%%%%%%%%%%%%%%%%%%%%%%%%%%%%%%%%%
\subsubsection{Benchmark for three-dimensional convection with modest
   viscosity variation}

\begin{figure}
 \unitlength 1mm
 \begin{picture}(140,30)
  \put(0,0){\framebox(140,30){Figure \ref{busse-surface}}}
 \end{picture}
 \caption{Isothermal surfaces obtained for the benchmark calculations
 of stationary convections in Busse et al.~\cite{busse93}. (a) Case 1a
 is for constant viscosity, while (b) Case 2 is for modestly
 temperature-dependent viscosity whose viscosity contrast is 20. The
 calculations were carried out with (a) $64\times32\times64$ and (b)
 $64\times64\times64$ mesh divisions.}
 \label{busse-surface}
\end{figure}

Second, we compare our results with those of Busse et al.~\cite{busse93}
for the cases of steady-state convection with both constant (Case 1a)
and modestly temperature-dependent viscosity (Case 2) in a
three-dimensional rectangular box of $a\times b\times 1$.
Figure \ref{busse-surface} shows the convective flow patterns for both
cases.
Case 1a is a bimodal convection of isoviscous fluid with
$Ra=3\times 10^4$ in a box of $a=1.0079$ and $b=0.6283$, while Case 2 is
a square-cell convection in a cube ($a=b=1$) with modestly
temperature-dependent viscosity.
In Case 2, the viscosity $\eta$ depends on temperature $T$ as,
\begin{equation}
 \begin{array}{l}
  \eta(T) = \exp\left[\bun{Q}{T+G}-\bun{Q}{0.5+G}\right],\\
  Q=\bun{225}{\ln(\reta)}-0.25\ln(\reta), G=\bun{15}{\ln(\reta)}-0.5.
 \end{array}
\end{equation}
Here, the viscosity contrast $\reta$ is set to be 20, and the Rayleigh
number defined by the viscosity for $T=0.5$ is $2\times10^4$.
In these cases the top and bottom surfaces are assumed to be rigid
($v_1=v_2=v_3=0$), while the vertical side walls are planes of mirror
symmetry.
The initial conditions are chosen to make single ascending and
descending flow at $(x,y)=(0,0)$ and $(a,b)$, respectively.

\begin{table}
 \caption{Benchmark comparison with Busse et al.~\cite{busse93}. See the
 text as for the meaning of the symbols.}
  \begin{center}
   \begin{tabular}{cccccc}
    \multicolumn{6}{l}{(a) Case 1a (constant viscosity, $Ra=3\times10^4$, $a=1.0079$, $b=0.6283$)} \\
    mesh & 16$\times$8$\times$16 & 32$\times$16$\times$32 &
    64$\times$32$\times$64 & 128$\times$64$\times$128& benchmark standard\\
    \hline
    $Nu$         &  3.7373 &  3.6019 &  3.5549 &  3.5419 & 3.5374$\pm$0.0005 \\
    $V_{\mathrm{rms}}$    &  42.047 &  41.382 &  41.104 &  41.026 & 40.999$\pm$0.004 \\
    $w(0,0,0.5)$ & 122.919 & 120.451 & 117.880 & 116.964 & 116.625$\pm$0.030 \\
    $w(0,b,0.5)$ &  20.044 &  33.693 &  38.566 &  39.994 & 40.500$\pm$0.030 \\
    $T(0,0,0.5)$ & 0.79161 & 0.80376 & 0.80262 & 0.80169 & 0.80130$\pm$0.00005 \\
    $T(0,b,0.5)$ & 0.57314 & 0.60324 & 0.61434 & 0.61761 & 0.61876$\pm$0.00005 \\
    $Q(0,0)$     & 7.68656 & 7.08517 & 6.82042 & 6.74082 & 6.7127$\pm$0.0500 \\
    $Q(a,0)$     & 1.85083 & 1.60300 & 1.53359 & 1.51460 & 1.5080$\pm$0.0500 \\
    $Q(0,b)$     & 2.75670 & 2.96542 & 3.10916 & 3.15671 & 3.1740$\pm$0.0500 \\
    $Q(a,b)$     & 0.74943 & 0.70104 & 0.70778 & 0.71225 & 0.7140$\pm$0.0500 \\
    \hline
    %\multicolumn{6}{c}{}\\
    \multicolumn{6}{l}{(b) Case 2 (modestly temperature-dependent
    viscosity, $a=b=1$)} \\
    mesh & 16$\times$16$\times$16 & 32$\times$32$\times$32 &
    64$\times$64$\times$64 & 128$\times$128$\times$128 & benchmark standard\\
    \hline
    $Nu$         &  3.0870 &  3.0503 &  3.0419 &  3.0399 & 3.0393$\pm$0.0050 \\
    $V_{\mathrm{rms}}$    &  35.350 &  35.161 &  35.130 &  35.124 & 35.13$\pm$0.05 \\
    $w(0,0,0.5)$ & 154.076 & 162.020 & 164.814 & 165.631 & 165.9$\pm$1.0 \\
    $w(0,b,0.5)$ & -25.662 & -26.446 & -26.650 & -26.703 & -26.72$\pm$0.1 \\
    $w(a,b,0.5)$ & -59.973 & -58.731 & -58.360 & -58.260 & -58.23$\pm$0.1 \\
    $T(0,0,0.5)$ & 0.88286 & 0.89803 & 0.90323 & 0.90474 & 0.90529$\pm$0.0010 \\
    $T(0,b,0.5)$ & 0.51736 & 0.50170 & 0.49724 & 0.49606 & 0.49565$\pm$0.0010 \\
    $T(a,b,0.5)$ & 0.25785 & 0.24417 & 0.24052 & 0.23957 & 0.23925$\pm$0.0010 \\
    $Q(0,0)$     & 6.09608 & 5.90091 & 5.85018 & 5.83783 & 5.834$\pm$0.015 \\
    $Q(a,0)$     & 1.80145 & 1.73607 & 1.71942 & 1.71508 & 1.714$\pm$0.015 \\
    $Q(a,b)$     & 0.78762 & 0.77281 & 0.76952 & 0.76869 & 0.768$\pm$0.015 \\
    \hline
   \end{tabular}
  \end{center}
 \label{busse-table}
\end{table}

We summarize the result of benchmark comparison for three-dimensional
convection in Table \ref{busse-table}.
We calculated the values of (i) $Nu$ and $V_{\mathrm{rms}}$, (ii)
vertical velocity $w$ and temperature $T$ at specified points at
mid-depth of the convecting vessel ($z=0.5$), and (iii) vertical
temperature gradient $Q$ at specified points at the top surface
$(z=0)$.
We also show in Table \ref{busse-table} the results from the benchmark
standard \cite{busse93}.
The agreement between the obtained results and the benchmark standard is
satisfactory for all of the values.
In particular, we obtained a good agreement for Case 2.
We thus conclude that the present numerical code accurately handles
three-dimensional convection of fluids with mildly variable viscosity.

%%%%%%%%%%%%%%%%%%%%%%%% Section 3.2.3 %%%%%%%%%%%%%%%%%%%%%%%%%%%%%%%%%%%%%%%%
\subsubsection{Benchmark for three-dimensional convection with strong
   viscosity variation}

\begin{table}
 \caption{Comparison of the flow patterns, Nusselt numbers $Nu$ and the
 minimum mesh sizes in $z$-direction $dz_{\mathrm{min}}$ between the
 present method (``KKS'') and Ogawa et al.~\cite{ogawa91}
 (``OSZ''). Note that in OSZ \cite{ogawa91} the mesh spacing was taken
 to be non-uniform with finer resolution near the boundaries so as to
 improve the accuracy in $Nu$.}
 \begin{center}
  \begin{tabular}{ccccccccc}
   Case & $Ra_t$ & $\reta$ & pattern & code &
   \multicolumn{3}{c}{KKS} & OSZ \\
   & & & & mesh &
   32$\times$16$\times$32 & 64$\times$32$\times$64 &
   128$\times$64$\times$128 &
   24$\times$14$\times$22$^{\ast}$ \\
   & & & & $dz_{\mathrm{min}}$ & 0.03125 & 0.015625 & 0.0078125 & $3.2\times10^{-2}$\\
   \hline
   1 & $10^5$ & $1$             & WL-3D &$Nu$& 9.816 & 10.071 & 10.101 & 9.72 \\
   4 & $10^3$ & $10^2$          & WL-2D &$Nu$& 4.060 &  4.073 &  4.077 & 4.13 \\
   10 & $10^3$ & $10^3$          & WL-3D &$Nu$& 4.961 &  4.932 &  4.921 & 4.96 \\
   \hline
   \\
   Case & $Ra_t$ & $\reta$ & pattern & code &
   \multicolumn{3}{c}{KKS} & OSZ \\
   & & & & mesh &
   32$\times$16$\times$32 & 64$\times$32$\times$64 &
   128$\times$64$\times$128 &
   44$\times$18$\times$30$^{\ast}$ \\
   & & & & $dz_{\mathrm{min}}$ & 0.03125 & 0.015625 & 0.0078125 & $1.7\times10^{-2}$\\
   \hline
   16 & $10^3$ & $3.2\times10^3$ & WL-3D &$Nu$& 5.284 & 5.271 & 5.271 & 5.37 \\
   \hline
   \\
   Case & $Ra_t$ & $\reta$ & pattern & code &
   \multicolumn{3}{c}{KKS} & OSZ \\
   & & & & mesh &
   32$\times$16$\times$32 & 64$\times$32$\times$64 &
   128$\times$64$\times$128 &
   44$\times$18$\times$30$^{\ast}$ \\
   & & & & $dz_{\mathrm{min}}$ & 0.03125 & 0.015625 & 0.0078125 & $1.3\times10^{-2}$\\
   \hline
   17 & $10^2$ & $3.2\times10^4$ & SL-3D &$Nu$& 3.561 & 3.631 & 3.659 & 3.61 \\
   18 & $32$   & $10^5$          & SL-3D &$Nu$& 3.119 & 3.174 & 3.197 & 3.17 \\
  \hline\end{tabular}
 \end{center}
 \label{ogawa-table}
\end{table}

\begin{figure}
 \unitlength 1mm
 \begin{picture}(140,30)
  \put(0,0){\framebox(140,30){Figure \ref{ogawa-surface}}}
 \end{picture}
 \caption{Isothermal surfaces for several cases listed in Table
 \ref{ogawa-table} obtained by three-dimensional convection with
 strongly temperature-dependent viscosity similar to Ogawa et
 al.~\cite{ogawa91}. The calculations were carried out with
 $64\times32\times64$ mesh divisions.}
 \label{ogawa-surface}
\end{figure}

We also carried out the calculations similar to Ogawa et
al.~\cite{ogawa91} in order to test our code for a much stronger
temperature-dependence of viscosity.
A steady-state convection in a box of $1.7\times0.5\times1$ is
considered.
The impermeable and shear-stress-free conditions are adopted along the
all boundaries.
We carried out calculations for several cases of Ogawa et
al.~\cite{ogawa91} with several mesh divisions.
In these calculations we varied $Ra_t$, the Rayleigh number defined with
$\eta_t$, and the viscosity contrast $\reta=\exp(E_T)$ between the top
and bottom boundaries, as listed in Table \ref{ogawa-table}.
Figure \ref{ogawa-surface} shows the convective flow patterns for
several cases.
We obtained the same flow patterns and the change in flow patterns
depending on $Ra_t$ and $\reta$ as in Ogawa et al.~\cite{ogawa91}.
The convective flow patterns are classified into three-dimensional
whole-layer convection (WL-3D) for Case 1 ($Ra_t=10^5$ and $\reta=1$),
two-dimensional roll of whole-layer convection (WL-2D) for Case 4
($Ra_t=10^3$ and $\reta=10^2$), again in WL-3D for Case 16 ($Ra_t=10^3$
and $\reta=3.2\times10^3$), and in three-dimensional stagnant-lid
convection (SL-3D) for Case 18 ($Ra_t=32$ and $\reta=10^5$).

We summarize the comparison of the flow patterns and Nusselt numbers
$Nu$ obtained by the present code (``KKS'') and those by Ogawa et
al.~\cite{ogawa91} (``OSZ'') in Table \ref{ogawa-table}.
The values of $Nu$ obtained by both codes agree within at most 4\%
deviation for all cases.
The largest deviation for Case 1 may come from the differences in
adopted mesh sizes.
The present calculations are carried out with the minimum mesh size in
$z$-direction of $1/128$, which is more than four times finer than that
employed in Ogawa et al.~\cite{ogawa91}.
It is most likely that a finer spatial resolution is required for Case 1
than that employed in Ogawa et al.~\cite{ogawa91} in order to resolve
the thermal boundary layers.

%%%%%%%%%%%%%%%%%%%%%%%%%%%% Section 3.3 %%%%%%%%%%%%%%%%%%%%%%%%%%%%%%%%%%%%%%
%\input{convergence.tex}
\subsection{Robustness and efficiency of multigrid iteration against
  spatial variation in viscosity}\label{ctest}

To demonstrate the robustness and efficiency of the present algorithm
against the spatial variation in viscosity, we performed convergence
tests similar to Albers \cite{albers00}.
Here we solve Eqs.~(\ref{mom_eq}) and (\ref{cont_eq}) only for
prescribed distributions of buoyancy and viscosity.
The distributions of buoyancy and viscosity are given by assuming the
Rayleigh number $Ra_t$ and the temperature-dependence of viscosity $E_T$
for a prescribed distribution of temperature.
(The distributions of temperature used for the tests will be introduced
below.)
We take into account the change of the viscosity variation in the
convecting vessel by increasing the global viscosity contrast
$\reta=\exp(E_T)$ from $1$ to $10^{10}$, and estimate a threshold value
of $\reta$ (hereafter denoted by $\reta^c$) below which the multigrid
iteration converges.
The values of velocity and pressure are initially set to zero, and are
iteratively updated until the $L^2$-norm of the residual
$\bm{r}=\bm{b}-\bm{A}\bm{x}$ in Eq.~(\ref{ax=b}) becomes smaller than
that of $\bm{b}$ by eight orders of magnitude.
All of the calculations were done with equally-spaced
$64\times64\times64$ mesh divisions and 6 grid levels, where the grid
is coarsest for the grid level $\ell=1$ and finest for $\ell=6$.
The mesh spacing is successively doubled as $\ell$ decreases, and the
mesh spacing is $1/2$ for grid level $\ell=1$.

\begin{figure}
 \unitlength 1mm
 \begin{picture}(140,30)
  \put(0,0){\framebox(140,30){Figure 3}}
 \end{picture}
 \caption{Distributions of temperature used for convergence tests. These
 temperature distributions were obtained by three-dimensional
 time-dependent simulation of thermal convection of fluids with
 temperature-dependent viscosity in a cubic box. The employed mesh
 division is $64\times64\times64$. In (a) the several isothermal
 surfaces are shown, while in (b) the horizontally-averaged temperature
 $T_h$ (black lines) and the maximum of $|grad(T)|$ at height $z$ (red
 lines) are plotted as a function of $z$.}
 \label{init-t}
\end{figure}

We used three distributions of temperature (hereafter denoted by
Temperatures A, B, and C).
In Figure \ref{init-t} we present (a), in the left column, several
isothermal surfaces and (b), in the right column, the plots of the
horizontally-averaged temperature $T_h$ and the maximum of the magnitude
of local temperature gradient $|grad(T)|_{\mathrm{max}}$ at each height
$z$, for these temperatures.
The temperature fields are taken from snapshots of a three-dimensional
time-dependent thermal convection of fluid with temperature-dependent
viscosity in a cubic box.
The Rayleigh number $Ra_t$ defined by viscosity $\eta_t$ is $10^3$, and
$E_T=\ln(10^5)$ is assumed.
In temperature A, which is taken from an early stage of the temporal
evolution (nondimensional time $t=5.4\times10^{-4}$), the convective
flow occurs only in the lowermost part of the box, and the local
variation in the temperature is very small in the entire box (see the
top panel of Figure \ref{init-t}b).
In temperature B ($t=1.52\times10^{-2}$), the cold fluid with $T<0.5$
sinks into the hot interior, and a large local variation in temperature
occurs around the descending flow near the bottom surface (see the
middle panel of Figure \ref{init-t}b).
In temperature C ($t=3.6\times10^{-2}$), the interior of the box is
nearly isothermal and, as can be also seen from the smaller
$|grad(T)|_{\mathrm{max}}$ than in Temperature B, the temperature
contrast between the descending flow and the surroundings becomes
smaller.

\begin{table}
 \caption{Numbers of pre- and post-smoothing calculations $\nsmooth$ on
 grid level $\ell$ employed in the various types of the smoothing
 procedures in this study. Here $N_{\mathrm{grid}}$ is the number of
 employed grids and $\reta$ is the magnitude of global viscosity
 contrast. In the present calculations $N_{\mathrm{grid}}$ is taken to
 be 6.}
 \begin{center}
  \begin{tabular}{c|c}
   Type & $\nsmooth(\ell)$ on grid level $\ell$\\
   \hline
   0 & $8$\\
   1 & $8\times2^{N_{\mathrm{grid}}-\ell}$\\
   %%% original %%%
   % 2 & $[\ln(r_\eta)-1]\times8\times2^{N_{\mathrm{grid}}-\ell}$\\
   % 3 & $[\ln(r_\eta)-1]\times8\times4^{N_{\mathrm{grid}}-\ell}$\\
   %%% revised %%%
   2 & $[\ln(r_\eta)+1]\times8\times2^{N_{\mathrm{grid}}-\ell}$\\
   3 & $[\ln(r_\eta)+1]\times8\times4^{N_{\mathrm{grid}}-\ell}$\\
   %%%
   \hline
  \end{tabular}
 \end{center}
 \label{nsmooth}
\end{table}

In addition, to find an appropriate implementation of the present
algorithm for the multigrid operation, we employed four different types
of smoothing procedures during the multigrid V-cycles.
In these types, we varied the number of iterations $\nsmooth$ by
(\ref{ax=b-iter}) for the pre- and post-smoothing calculations at each
grid level, as listed in Table \ref{nsmooth}.
In Type 0, $\nsmooth$ is taken to be 8 at all grid levels, and is the
smallest among the four types employed in these tests.
In Type 1, $\nsmooth$ depends on the grid level $\ell$.
The value of $\nsmooth$ is taken to be equal to that in Type 0 at the
finest grid ($\ell=6$), and is successively doubled as the grid becomes
coarser.
In Types 2 and 3, in contrast, $\nsmooth$ is assumed to be dependent on
both the grid level $\ell$ and the global viscosity contrast $\reta$.
In both types, $\nsmooth$ at the finest grid is taken to be
$[\ln(\reta)+1]$ times larger than that in Type 1, and the values of
$\nsmooth$ is successively multiplied by two and four as the grid becomes
coarser in Types 2 and 3, respectively.

\begin{figure}
 \unitlength 1mm
 \begin{picture}(140,30)
  \put(0,0){\framebox(140,30){Figure 4}}
 \end{picture}
 \caption{Convergence behavior for various values of global viscosity
 contrast $\reta$ and for temperature fields in Figure \ref{init-t} by
 using various types of smoothing procedures listed in Table
 \ref{nsmooth}. In (a) the number of V-cycles $\nvcycle$ is plotted
 against $\reta$. In (b) the measure of total computational cost in the
 smoothing procedures $\wsmooth$ is plotted against $\reta$. The values
 are plotted only for $\reta\leq\reta^c$, below which the multigrid
 iteration converges.}
 \label{conv-graph}
\end{figure}

We compared the convergence behaviors of different types of smoothing
procedures, for three distributions of $T$ and various values of
$\reta$.
In Figure \ref{conv-graph} we show the plots against $\reta$ of (a) the
number of V-cycles $\nvcycle$ and (b) the function $\wsmooth$ given by,
\begin{equation}
 \wsmooth = \bun{\sum_{\ell=1}^{6}n(\ell)\times m(\ell)}{m(\ell=6)},
  \label{def-wsmooth}
\end{equation}
where $n(\ell)$ is the total number of iterations by (\ref{ax=b-iter})
in the smoothing procedures on grid level $\ell$, and
$m(\ell)\equiv4\times(64/2^{6-\ell})^3$ is the number of elements of the
unknown vector $\bm{x}$ (i.e., the number of unknown variables; see
(\ref{ax=b-iter})) on grid level $\ell$.
The numerator in (\ref{def-wsmooth}) represents the total number of
updating operations of unknown variables during the entire multigrid
iterations, while the denominator represents the number of updating
operation during one smoothing calculation at the finest grid
($\ell=6$).
Namely, $\wsmooth$ is the measure of the computational cost of the
multigrid iterations, and it indicates that the total computational cost
spent in the smoothing procedures during the entire multigrid iterations
is equal to that virtually spent in $\wsmooth$-times smoothing
calculations on the finest grid.
In the figure, the values are plotted only for $\reta\leq\reta^c$, below
which the multigrid iteration converges.

The comparisons of $\reta^c$ for different types of smoothing procedures
in Figure \ref{conv-graph} show that the value of $\reta^c$ becomes
larger from Types 0 to 3 for all temperature distributions.
For Type 0, $\reta^c$ is $10^3$ for Temperature A, and $10^2$ for
Temperatures B and C.
By changing from Types 0 to 1, $\reta^c$ increases to $10^7$ for
Temperature A, and to $10^3$ for Temperatures B and C.
The value of $\reta^c$ significantly increases by further changing to
Types 2 and 3.
For Type 2, $\reta^c$ is larger than $10^{10}$ for Temperature A, and
$10^6$ for Temperatures B and C.
For Type 3, $\reta^c$ is larger than $10^{10}$ for Temperature A, and
$10^8$ for Temperatures B and C.
From the comparison between these types, we conclude that the robustness
of the multigrid iteration is improved by increasing $\nsmooth$.
In particular, a significant improvement is obtained for large $\reta$
by increasing $\nsmooth$ in proportion to $[\ln(\reta)+1]$.

\begin{figure}
 \unitlength 1mm
 \begin{picture}(140,30)
  \put(0,0){\framebox(140,30){Figure 5}}
 \end{picture}
 \caption{Comparison in the evolution of the $L^2$-norm of residuals on
 the finest grid (grid level $\ell=6$) during the initial five multigrid
 V-cycles between the case with the smoothing procedures of (a) Type 1
 and (b) Type 0. Both calculations were performed with Temperature A and
 $\reta=10^4$. The red lines indicate the evolutions obtained by the
 smoothing calculations on the grid level $\ell=6$, while the blue
 arrows indicate the changes of residuals for $\ell=6$ due to the
 coarse-grid correction.}
 \label{conv-or-not}
\end{figure}

The multigrid iteration becomes more robust against $\reta$ for larger
$\nsmooth$, since larger number of iterations of (\ref{ax=b-iter}) can
reduce the errors more effectively.
To see this more clearly, we show in Figure \ref{conv-or-not} the
evolution of the $L^2$-norm of residual during the multigrid iteration
on the finest grid ($\ell=6$) for the cases (a) where the iteration
converged with Type 1, and (b) where the iteration diverged with Type 0.
For Type 1 where the multigrid iteration converges (Figure
\ref{conv-or-not}a), the $L^2$-norm of residual on the finest grid is
significantly reduced after coming back from the smoothing on coarser
grids.
Even if the residual is increased during the smoothing calculations on
the finest grid (it can sometimes happen in this kind of decaying
oscillation system; see eq.~(\ref{divv_evol}) or
(\ref{divv_evol_variable_eta})), it is significantly reduced by the
coarse-grid correction.
As a result, the residual is successively reduced by the multigrid
iterations.
For Type 0, where the multigrid iteration diverges (Figure
\ref{conv-or-not}b), in contrast, the $L^2$-norm of residual on the
finest grid is increased after coming back from the smoothing on coarser
grids.
The smoothing calculations on the finest grid reduces the residual,
except for the first multigrid iteration.
However, the decrease in the residual by the smoothing calculations is
too small to overcome the increase by the coarse-grid correction.
As a result, the residual is successively increased by the multigrid
iterations.
From the comparison between these cases, we conclude that sufficient
number of smoothing calculations are necessary on coarser grids in order
to ensure the convergence of multigrid iteration when a large spatial
variation in viscosity is involved.

Figure \ref{conv-graph} also shows that the convergence rates of our
method differ between the types of smoothing procedures.
The comparison in the convergence rates between Types 0 and 1 for small
$\reta$ indicates that a successive increase in $\nsmooth$ to coarser
grids significantly reduces the number of V-cycles $\nvcycle$ (Figure
\ref{conv-graph}a) as well as the computational cost $\wsmooth$ (Figure
\ref{conv-graph}b).
This stabilizing effect of increasing the number of smoothing iterations
on coarser grids is in agreement with the features of previous results
using multigrid methods \cite[for example]{trompert96,albers00}.
The comparison between Types 1 to 3 in Figure \ref{conv-graph}a indicate
that $\nvcycle$ is significantly reduced by increasing $\nsmooth$ in
proportion to $[\ln(\reta)+1]$.
This feature is also consistent with the earlier results \cite[for
example]{trompert96,albers00}.
As can be seen in Figure \ref{conv-graph}b, however, the computational
cost $\wsmooth$ of smoothing calculations increases as $\nsmooth$
becomes larger from Types 1 to 3.
This is because the cost of smoothing calculations during one V-cycle
becomes significantly larger.

We also note from Figure \ref{conv-graph} that the convergence behavior
differs between the prescribed temperature fields.
This difference may reflect the difference in the magnitude of the local
temperature gradient $|grad(T)|$ shown in Figure \ref{init-t}b.
The different temperature distributions provide the different
distributions of viscosity.
The distribution of viscosity determines the nature of the coefficient
matrix $\bm{A}$ in (\ref{ax=b}) and, hence, determines the convergence
behavior of (\ref{ax=b-iter}).
In particular, a larger $|grad(T)|$ generates a larger local variation
of viscosity for given $\reta$, and makes the convergence of
(\ref{ax=b-iter}) more difficult.
The convergence behaviors presented in Figure \ref{conv-graph} are
qualitatively consistent with the above conjecture.
The multigrid iteration converges most easily for Temperature A with the
smallest $|grad(T)|$, and least easily for Temperature B with the
largest $|grad(T)|$.
In Figure \ref{conv-graph}, the value of $\reta^c$ is the largest for
Temperature A for any type of smoothing procedures.
Although the values of $\reta^c$ are the same for Temperatures B and C
when the same type of smoothing procedures is employed, the number of
V-cycles $\nvcycle$ is smaller for Temperature C than for Temperature
B, except for the case with Type 2 for Temperature C and $\reta=10^6$.
These convergence behaviors are also consistent with those in the
earlier studies \cite[for example]{trompert98a,albers00}.

%%%%%%%%%%%%%%%%%%%%%%%%%%%% Section 4 %%%%%%%%%%%%%%%%%%%%%%%%%%%%%%%%%%%%%%%%
%\input{discus.tex}
\section{Discussion and concluding remarks}

We developed a numerical algorithm for solving mantle convection
problems with strongly variable viscosity.
Equations for conservation of mass and momentum for highly viscous and
incompressible fluids are solved iteratively by a multigrid method in
combination with pseudo-compressibility and local time stepping
techniques.
In order to demonstrate its efficiency, the present algorithm has been
implemented into a mantle convection simulation program based on the
finite-volume discretization in a three-dimensional rectangular domain.
Benchmark comparison with previous two- and three-dimensional
calculations including the temperature- and/or depth-dependent
viscosity revealed that accurate results are obtained even for the
cases with viscosity variations of several orders of magnitude.
We could also significantly improve the robustness of the numerical
method against a spatial variation in viscosity by increasing the pre-
and post-smoothing calculations in the multigrid operations.
%%% original %%%
%We could simulated the cases even for the viscosity contrasts up to
%$10^{10}$, although the convergence rate deteriorates with increasing
%viscosity variations.
%%% revised %%%
We achieved convergence even for the viscosity contrasts up to
$10^{10}$, although the convergence rate deteriorates with increasing
viscosity variations.
%%%
The present algorithm can be further applied to the numerical models
under more realistic conditions, such as spherical shell geometry and so
on.

The results of convergence tests described in Section \ref{ctest}
suggest that the convergence of multigrid method in combination with the
present algorithm is determined by the accuracy of the coarse-grid
correction when a large spatial variation in viscosity is incorporated.
In the present convergence tests, the accuracy of coarse-grid correction
was improved by increasing the amount of smoothing calculations on
coarser grids during one multigrid iteration.
This strategy always works, since the present iterative procedure never
diverges as long as the local time stepping satisfies the CFL condition
(see Section \ref{lts}).
However, it turned out that total computational costs significantly
increased as the magnitude of viscosity contrast becomes larger (see
Figure \ref{conv-graph}b).
Therefore, one needs to further accelerate the rate convergence of
(\ref{ax=b-iter}) in order to improve the efficiency of smoothing
calculations and, in turn, to improve the accuracy of coarse-grid
correction.

The discussion in Section \ref{lts} also suggests that the rate of
convergence of the present smoothing algorithm can be accelerated by
choosing the matrix $\bm{T}$ in (\ref{ax=b-iter}) more properly.
%%% original %%%
%As has been demonstrated in Section \ref{lts}, the convergence of
%(\ref{ax=b-iter}) becomes faster as the spectral radius of the matrix
%$\bm{I}-\bm{T}\bm{A}$ (or, equivalently, $\bm{I}-\bm{A}\bm{T}$) becomes
%smaller.
%%% revised %%%
As has been pointed out in Section \ref{lts}, the matrix $\bm{T}$ acts
as a ``preconditioner'' for (\ref{ax=b}).
That is, the convergence of (\ref{ax=b-iter}) becomes faster as $\bm{T}$
better approximates $\bm{A}^{-1}$.
%%%
In the present application to mantle convection problems, we had chosen
$\bm{T}$ as a diagonal matrix by the use of the local time stepping
approach.
In addition, we had determined the values of nonzero elements of
$\bm{T}$ from the assumption that both the rate of ``diffusion'' in
velocity components and the rate of propagation of ``pseudo-sound wave''
were kept almost uniform in space.
%%% revised
This choice of $\bm{T}$ corresponds to the preconditioning similar to
the diagonal scaling of (\ref{ax=b}).
%%%
%%% original %%%
%However, apart from the local time stepping approach, we can use any
%arbitrary matrix $\bm{T}$, rather than a diagonal matrix, as long as
%$\bm{T}$ properly approximates $\bm{A}^{-1}$.
%%% revised %%%
However, apart from the local time stepping approach, we can use any
arbitrary matrix $\bm{T}$ rather than a diagonal matrix.
In other words, the convergence of (\ref{ax=b-iter}) is
% most likely to be
accelerated if $\bm{T}$ preconditions (\ref{ax=b}) more properly than
in the present paper.
%%%
We can construct $\bm{T}$, for example, by an incomplete factorization
of $\bm{A}$.
Since the matrix $\bm{T}$ defined thus is most likely to be a better
approximation of $\bm{A}^{-1}$ than that employed here, the convergence
of (\ref{ax=b-iter}) is expected to be faster than in the present
cases.
On the other hand, constructing $\bm{T}$ by an incomplete factorization
has several disadvantages because (i) more memory is required to store
all of the nonzero elements of $\bm{T}$ and (ii) the operation of an
incomplete factorization is difficult to vectorize and parallelize.
%%% original %%%
%We should, therefore, take into account the employed computer
%architecture (such as scalar or vector processors) in choosing the
%most appropriate $\bm{T}$.
%%% revised %%%
We should, therefore, take into account the specific computer
architecture (such as scalar or vector processors) in choosing the
most appropriate $\bm{T}$ in terms of preconditioning techniques.
%%%

However, the most essential improvement possible for the present
numerical method is the multigrid operation itself, rather than the
smoothing calculations during multigrid iteration.
As has been suggested in Section \ref{ctest}, the reason why the
multigrid iteration diverges for larger $\reta$ is that the coarse-grid
correction becomes less efficient as $\reta$ increases (see Figures
\ref{conv-graph} and \ref{conv-or-not}).
The inefficiency of coarse-grid correction may come from an
inappropriate treatment of the influences of variable viscosity in the
multigrid operation.
In the present numerical method, as well as in most of earlier models
based on finite-volume discretization
\cite{tackley-thesis,trompert96,albers00}, we followed a ``standard''
strategy of multigrid.
Namely, a linear interpolation was used for transferring the residual
and error between adjacent grid levels.
In addition, the discretized equations on coarser grids were derived by
directly discretizing the differential equations on the particular
coarser grids.
However, as has been already acknowledged in the literature of multigrid
\cite{wesseling-book,trottenberg01}, the standard strategy does not
efficiently work when the differential equations contain strongly
varying coefficients.
One of the potential remedies is to develop the discretized equations
based on Galerkin coarse-grid approximation
\cite{wesseling-book,trottenberg01}, where the coefficient matrix
$\bar{\bm{A}}$ on a coarse grid is defined from the matrix on a fine
grid $\bm{A}$ by,
\begin{equation}
 \bar{\bm{A}} = \bm{R}\bm{A}\bm{P},
\end{equation}
where $\bm{R}$ and $\bm{P}$ are the restriction and prolongation
operators, respectively.
The Galerkin coarse-grid approximation can automatically define the
coefficient matrix $\bar{\bm{A}}$ on the coarse grid, so as to
sufficiently approximate $\bm{A}$ defined on the fine grid.
But it is not straightforward to use this approach in finite-volume
models, since it makes the coefficient matrices on the coarse grids very
complicated, with increased stencils, compared to those coming from
the direct discretization.
On the other hand, another potential remedy is to use the
operator-dependent transfer operators
\cite{wesseling-book,trottenberg01}.
These transfer operators utilize the structure of $\bm{A}$ with the
variations in its coefficients and, hence, improve the accuracy of the
restriction and prolongation operations.
Indeed, Yang and Baumgardner \cite{yang-baum00} incorporated these
transfer operators together with the Galerkin coarse-grid operators into
their finite-element models, and demonstrated significant improvements
in the robustness and efficiency of their multigrid procedures even for
the cases with strongly viscosity variations.
We expect that the robustness of our multigrid method presented in
Section \ref{result-section} can be improved by incorporating the
operator-dependent transfer operators into the present model, although
their efficiency is left uncertain when used without the Galerkin
approximation \cite{albers00}.
%%%

It is also an important issue to improve the robustness of the multigrid
iterations against a ``sharp'' or ``discontinuous'' variation in
viscosity, in addition to a ``smooth'' variation considered here.
One of the major unsolved problems in the numerical study of mantle
dynamics is to reproduce the motion of surface plates in the framework
of mantle convection.
Earlier numerical studies \cite[for a review]{bercovici03a} had
demonstrated that the generation of localized zones of low viscosity
around the plate boundaries plays an important role in reconciling the
fluid-like flow of mantle with the discrete motion of surface plates.
However, it is very difficult in essence for the multigrid method to
deal with a local variation in viscosity as long as uniform mesh spacing
is considered, since such local features are hardly ``visible'' on
coarser grids.
%%% original %%%
%Recently, on the other hand, Albers \cite{albers00} successfully applied
%the technique of local mesh refinement to a multigrid method for mantle
%convection with a strong variation in viscosity.
%Since this method allows coarsening the grids non-uniformly in space, a
%local variation of viscosity can be handled accurately by, for example,
%not coarsening the grids around the regions of local variation.
%We thus speculate that the employment of the local mesh refinement
%technique together with the present method is one promising approach to
%reproduce the motion of surface plates in the numerical model of mantle
%convection.
%%% revised %%%
Recently, in several fields of geophysical fluid dynamics, new numerical
techniques, such as the local mesh refinement \cite{albers00}, its
combination with spectral-element method \cite{fournier04}, and the
wavelet-based method \cite{vasilyev97}, are utilized.
Since these methods allow taking the spatial resolution non-uniformly in
space, a local variation of viscosity can be handled accurately by, for
example, using finer resolution around the regions of local variation.
We thus speculate that the combination of these techniques together with
the present method is one promising approach to reproduce the motion of
surface plates in the numerical model of mantle convection.
%%%

%%%%%%%%%%%%%%%%%%%%%%%%%% Acknowledgement %%%%%%%%%%%%%%%%%%%%%%%%%%%%%%%%%%%%
%\input{shaji.tex}
\section*{Acknowledgement}
We thank Tomoeki Nakakuki, David A.\ Yuen, Paul J.\ Tackley, and Masaki
Ogawa for discussion and comments.
We also thank two anonymous reviewers for valuable comments which
greatly improved the manuscript.
The calculations presented in this paper were in part done by Earth
Simulator at Japan Agency for Marine-Earth Science and Technology.

%%%%%%%%%%%%%%%%%%%%%%%%%%% References %%%%%%%%%%%%%%%%%%%%%%%%%%%%%%%%%%%%%%%%
%\clearpage
\bibliographystyle{unsrt}
%\bibliography{localref}

\clearpage

%%%%%%%%%%%%%%%%%%%%%%%%%%% Figures and Captions %%%%%%%%%%%%%%%%%%%%%%%%%%%%%%
%\input{figures.tex}
%\clearpage
\begin{center}
 Figure \ref{busse-surface}

 \vspace{\baselineskip}
 \setlength{\thisheight}{0.33\textheight}
 \begin{tabular}{cc}
  (a) Case 1a & (b) Case 2 \\
  \includegraphics[height=\thisheight,keepaspectratio,clip]{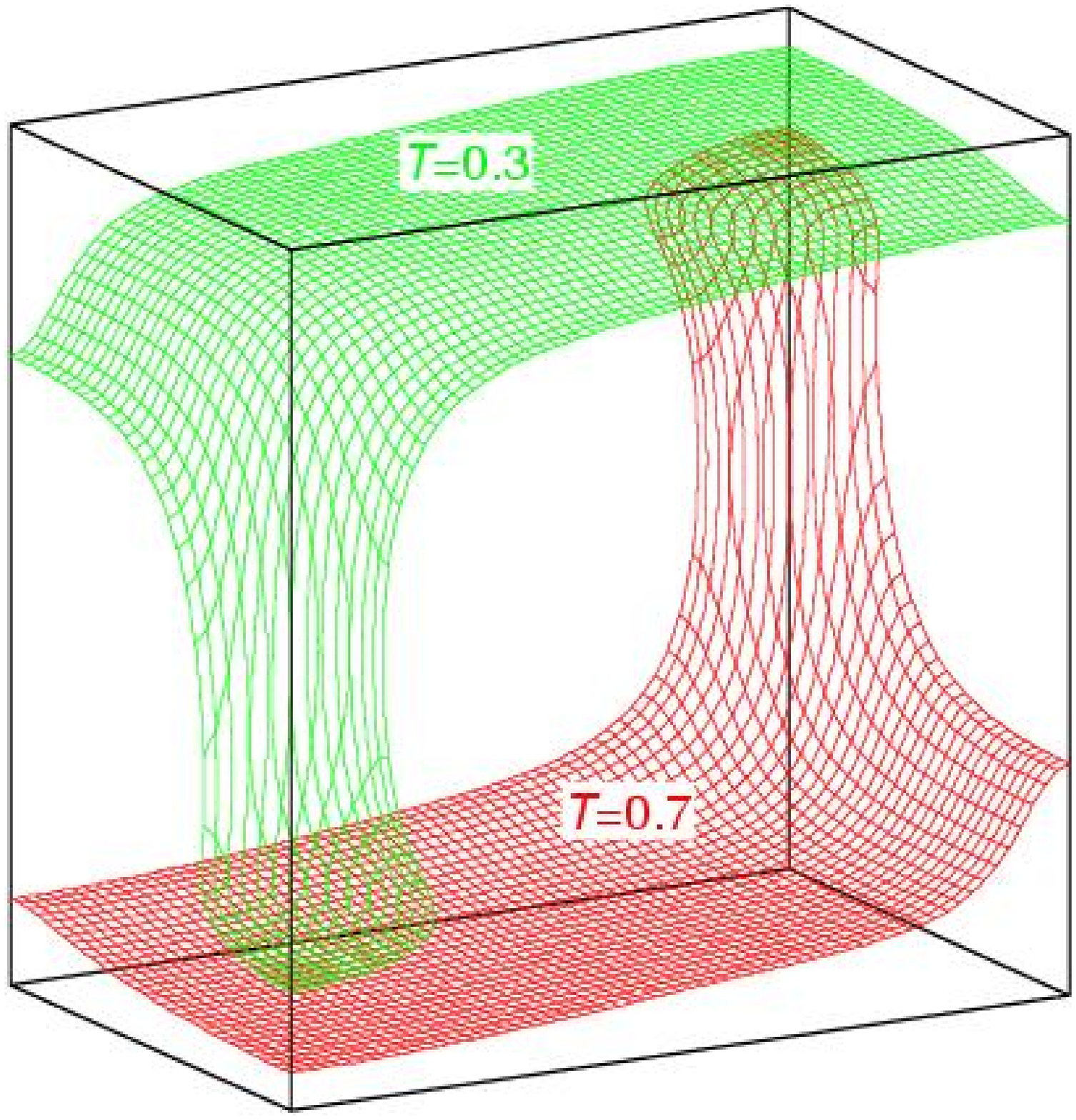}
  &
  \includegraphics[height=\thisheight,keepaspectratio,clip]{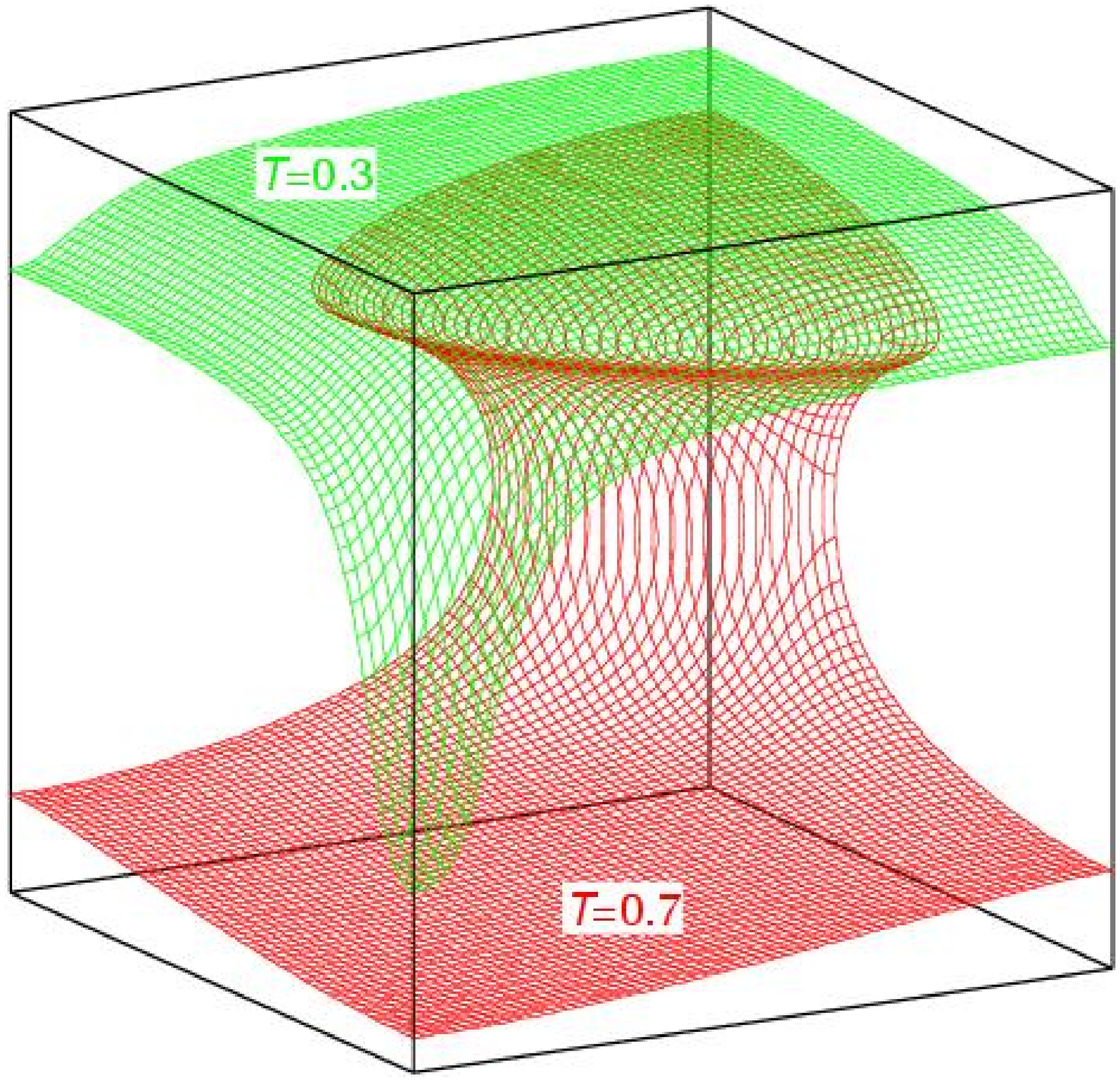}
  \\
 \end{tabular}
\end{center}

\clearpage
\begin{center}
 Figure \ref{ogawa-surface}

 \vspace{\baselineskip}
 \setlength{\thiswidth}{0.5\textwidth}
 \begin{tabular}{cc}
  (a) Case 1 ($Ra_t=10^5$, $\reta=1$) &
  (b) Case 4 ($Ra_t=10^3$, $\reta=10^2$) \\
  \includegraphics[width=\thiswidth,keepaspectratio,clip]{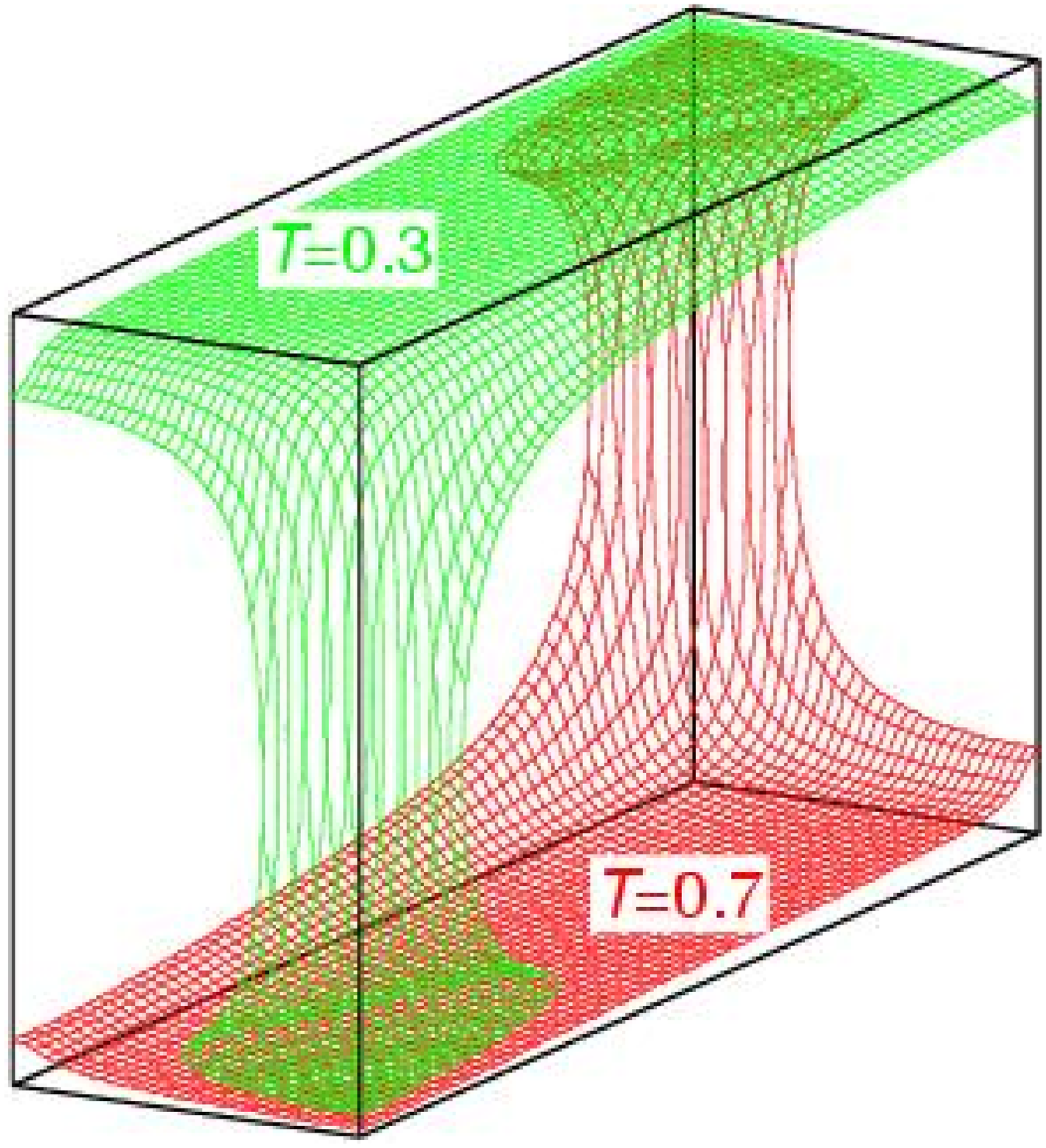} &
  \includegraphics[width=\thiswidth,keepaspectratio,clip]{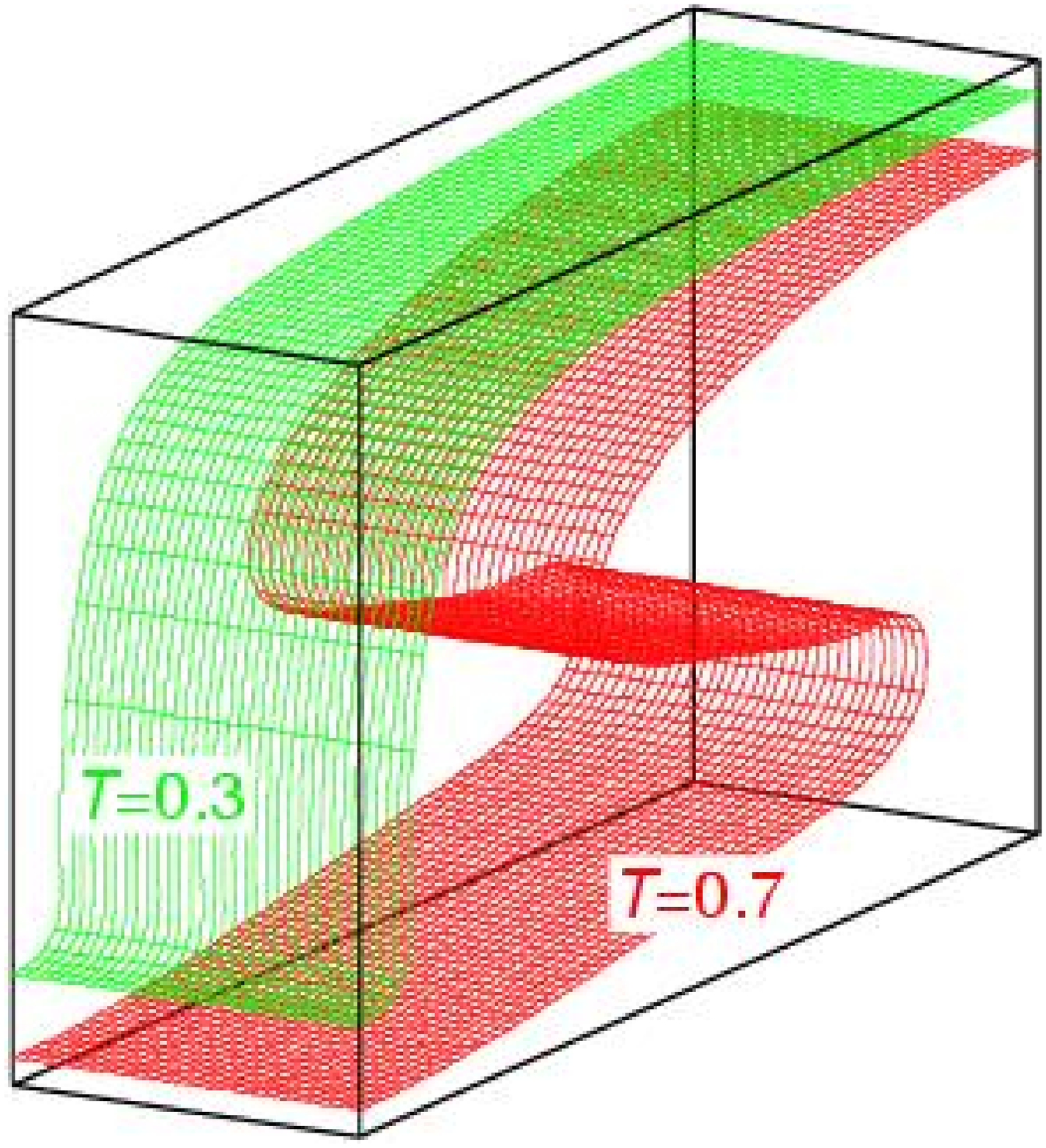} \\
  (c) Case 16 ($Ra_t=10^3$, $\reta=3.2\times10^3$) &
 (d) Case 18 ($Ra_t=32$, $\reta=10^5$) \\
  \includegraphics[width=\thiswidth,keepaspectratio,clip]{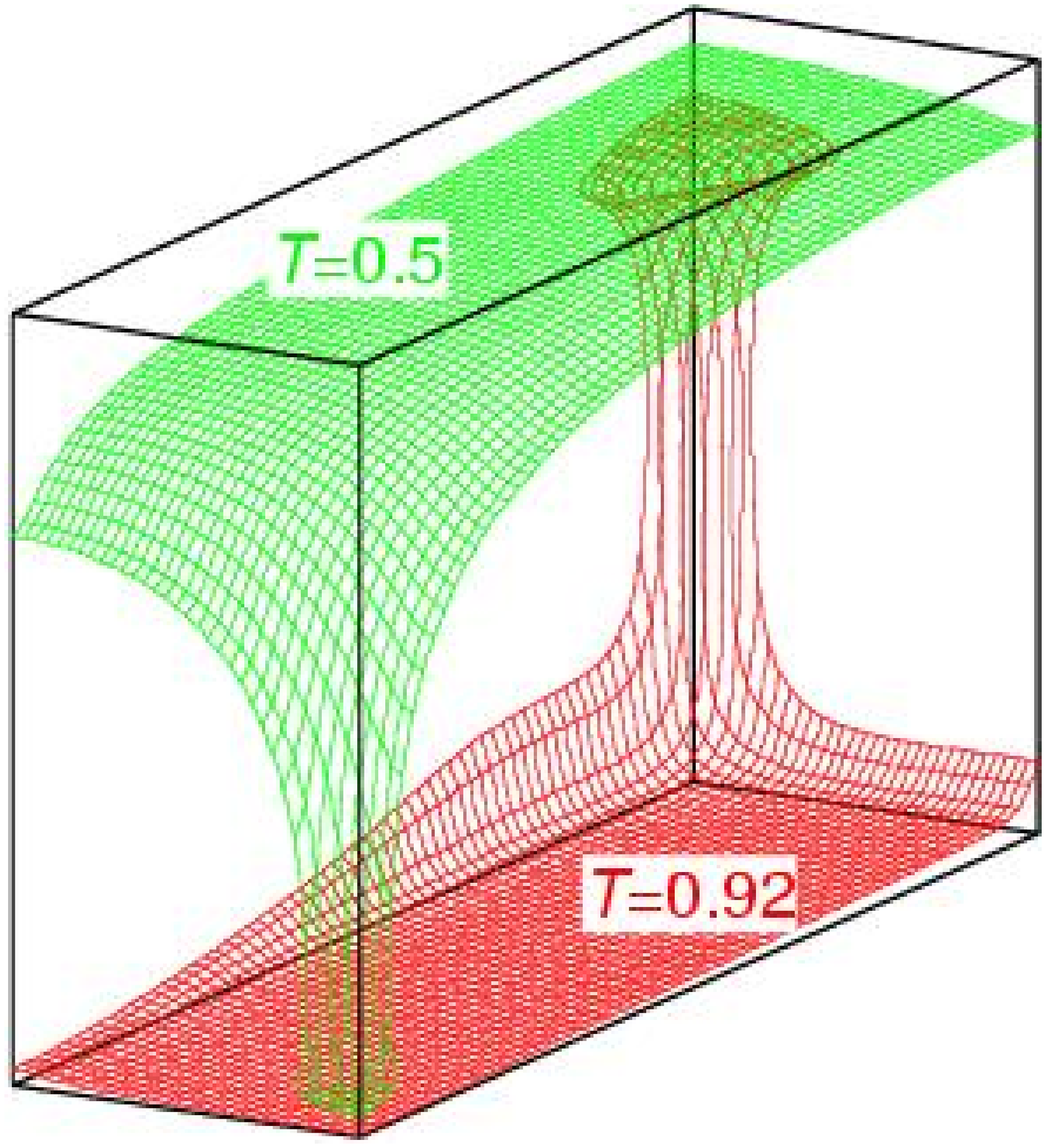} &
  \includegraphics[width=\thiswidth,keepaspectratio,clip]{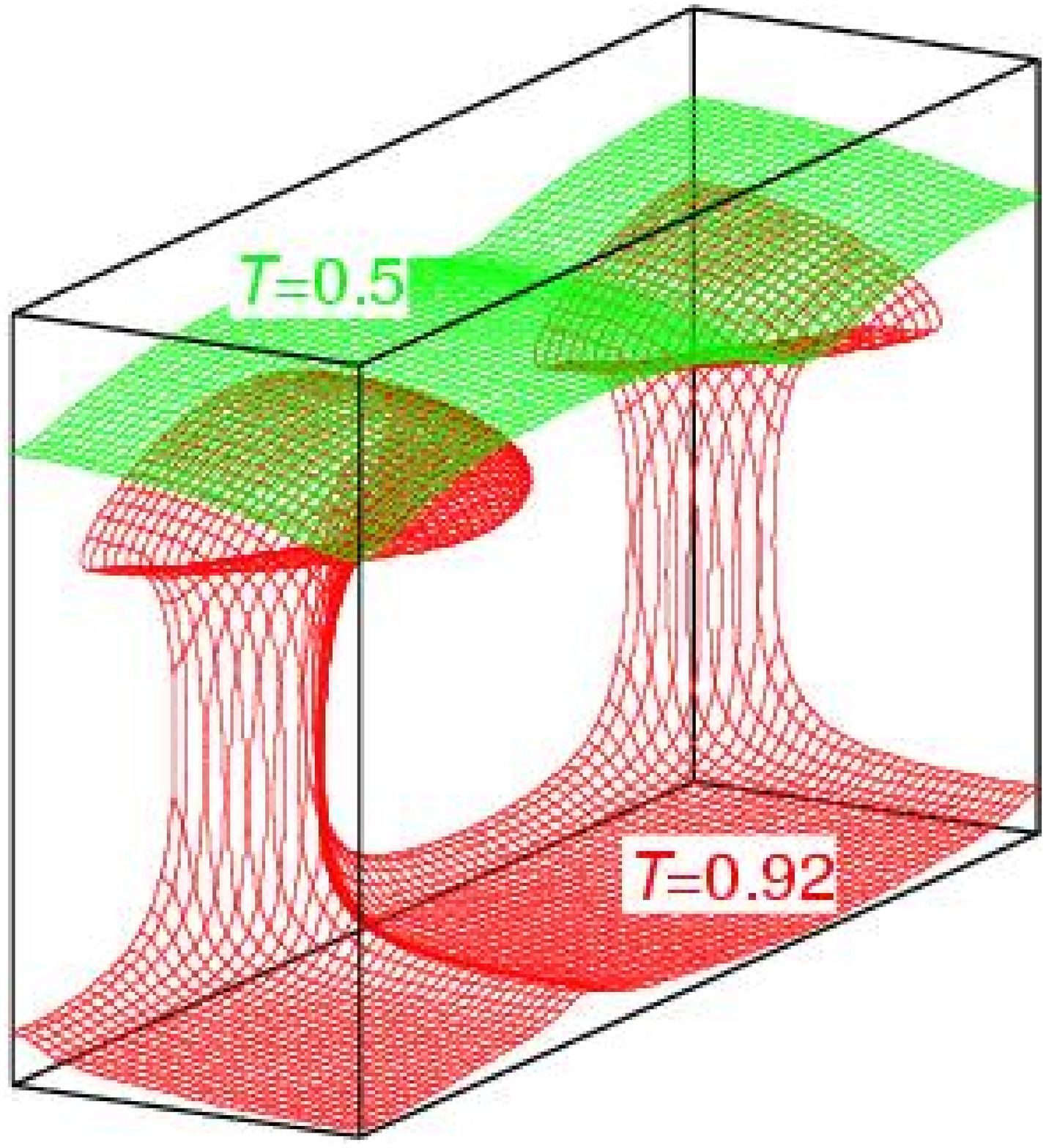} \\
 \end{tabular}
\end{center}

\clearpage
\begin{center}
 Figure \ref{init-t}

 \vspace{\baselineskip}
 \setlength{\thisheight}{0.26\textheight}
 \begin{tabular}{cc}
  (a) isotherms & (b) $T_h(z)$ and $|grad(T)|_{\mathrm{max}}(z)$ \\
  %\hline
  \multicolumn{2}{c}{Temperature A ($t=5.4\times10^{-4}$)}\\
  \includegraphics[height=\thisheight,keepaspectratio,clip]{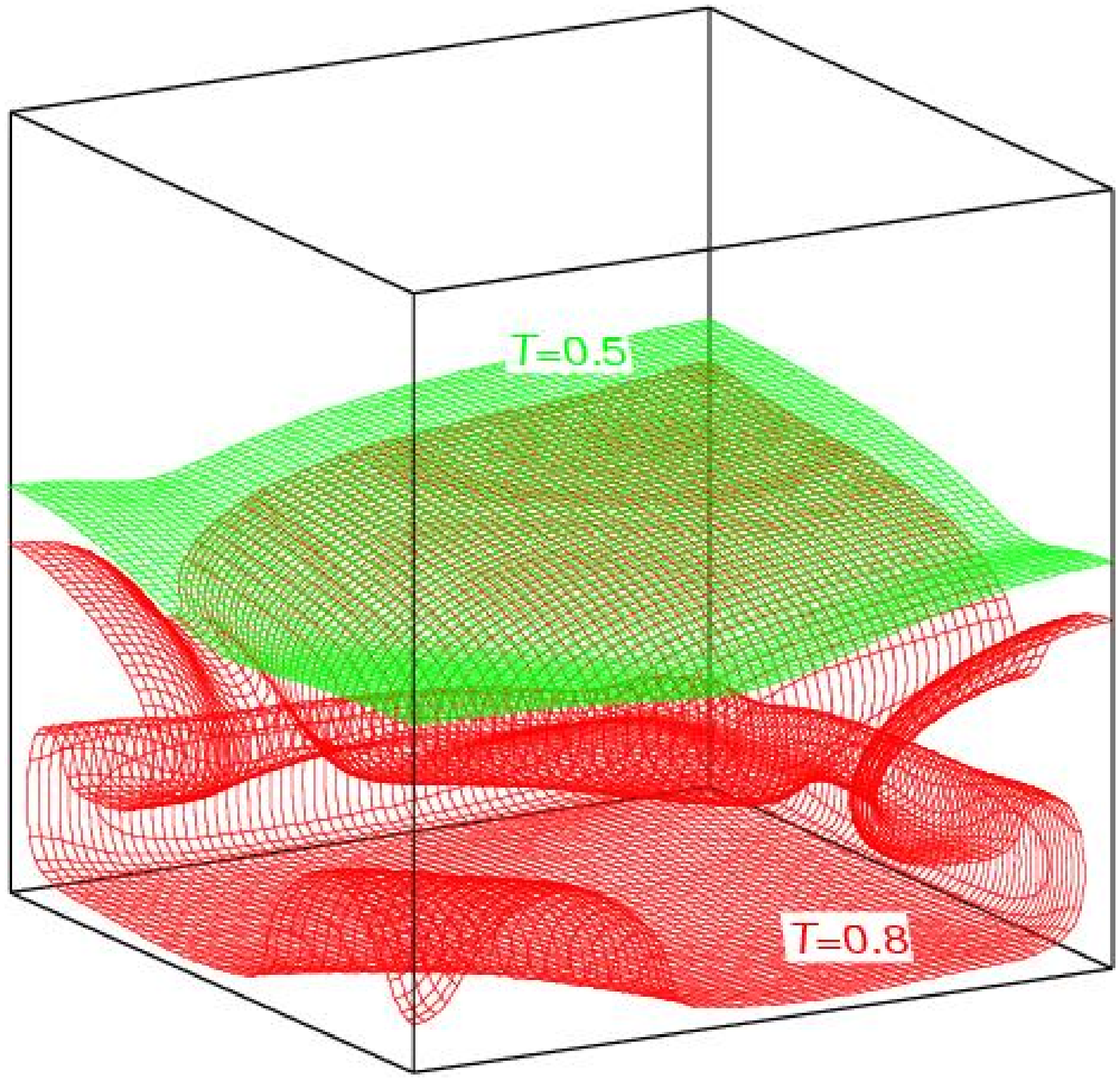}
  &
  \includegraphics[height=\thisheight,keepaspectratio,clip]{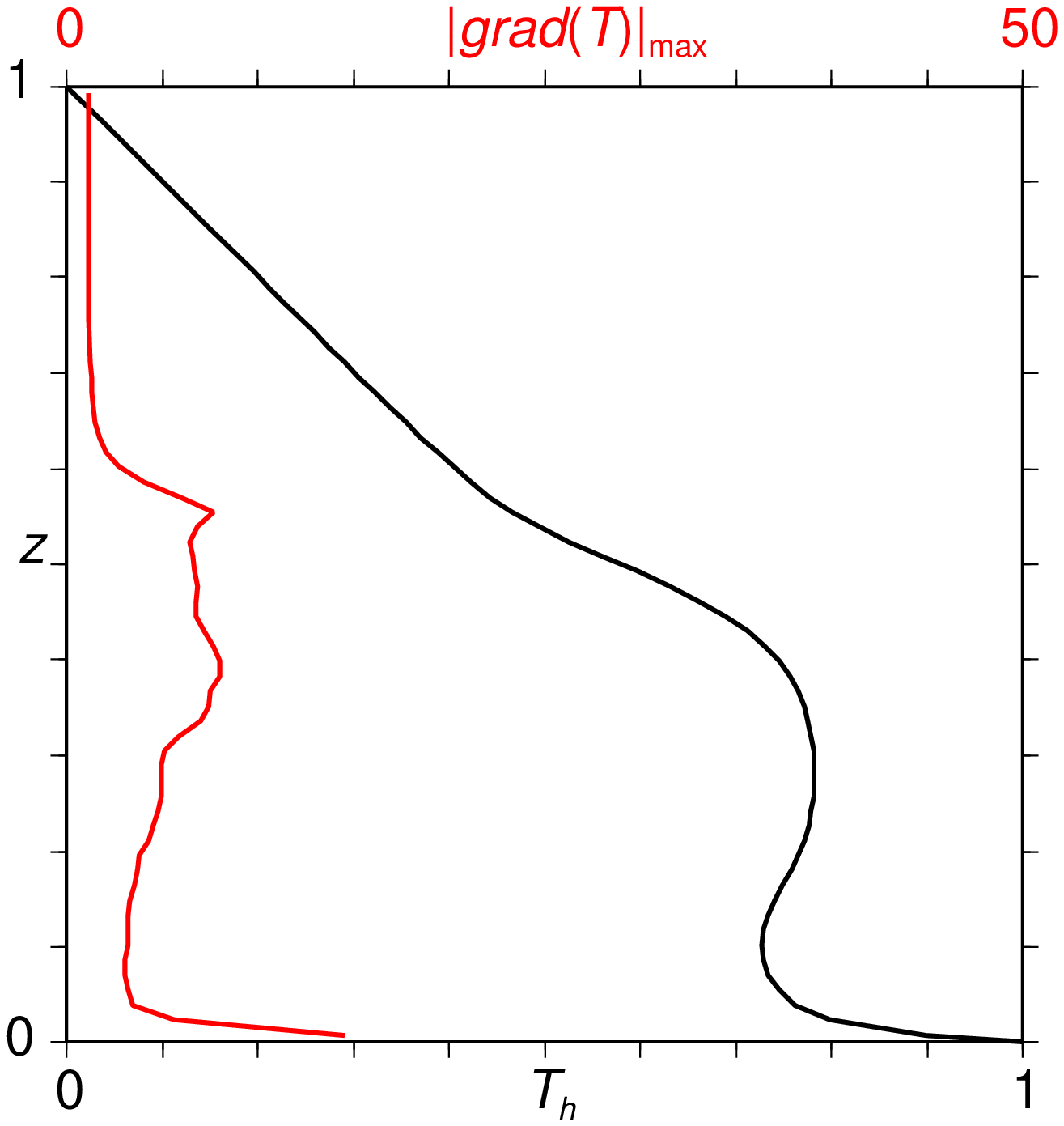}
  \\
  %\hline
  \multicolumn{2}{c}{Temperature B ($t=1.52\times10^{-2}$)} \\
  \includegraphics[height=\thisheight,keepaspectratio,clip]{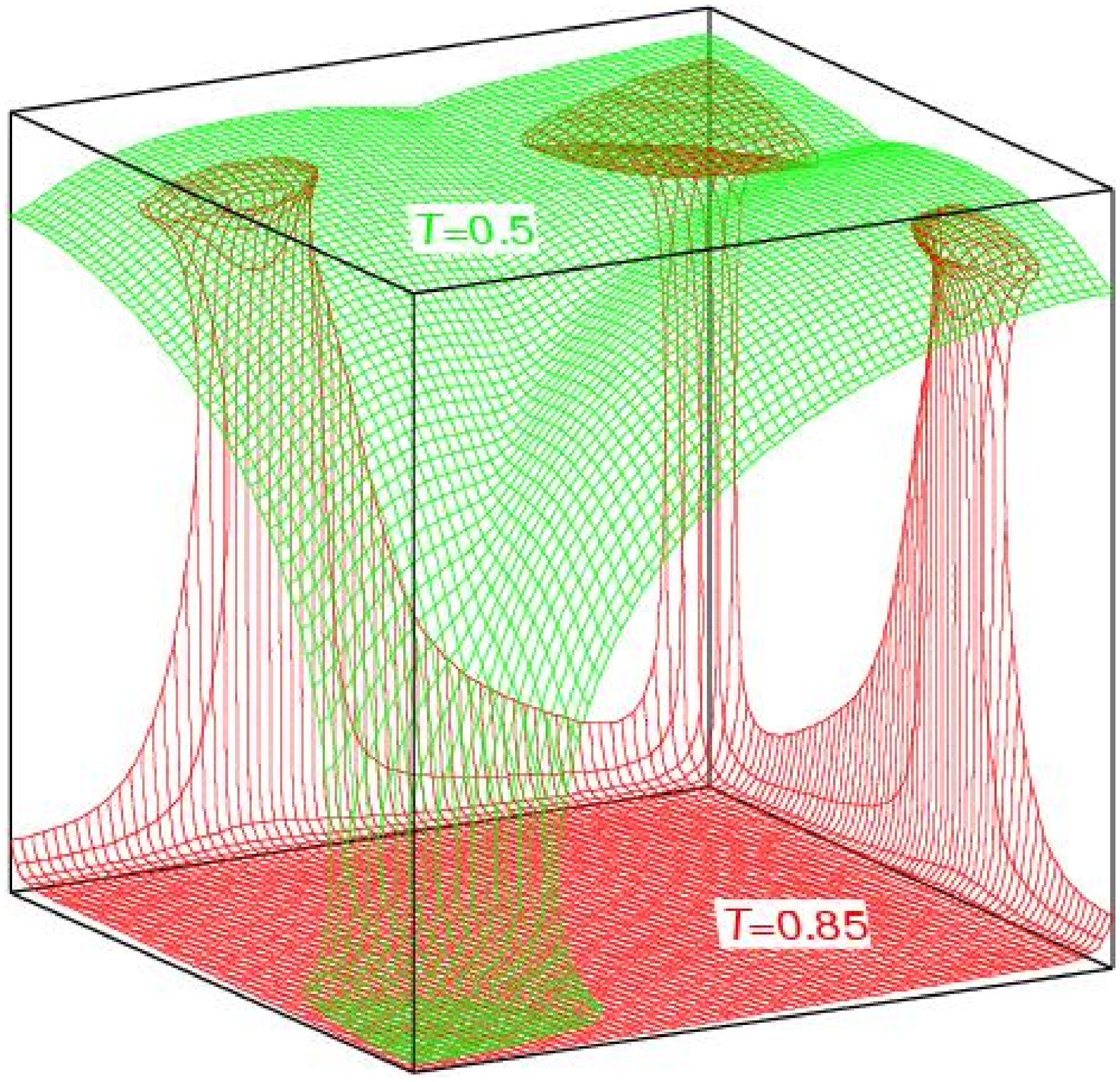}
  &
  \includegraphics[height=\thisheight,keepaspectratio,clip]{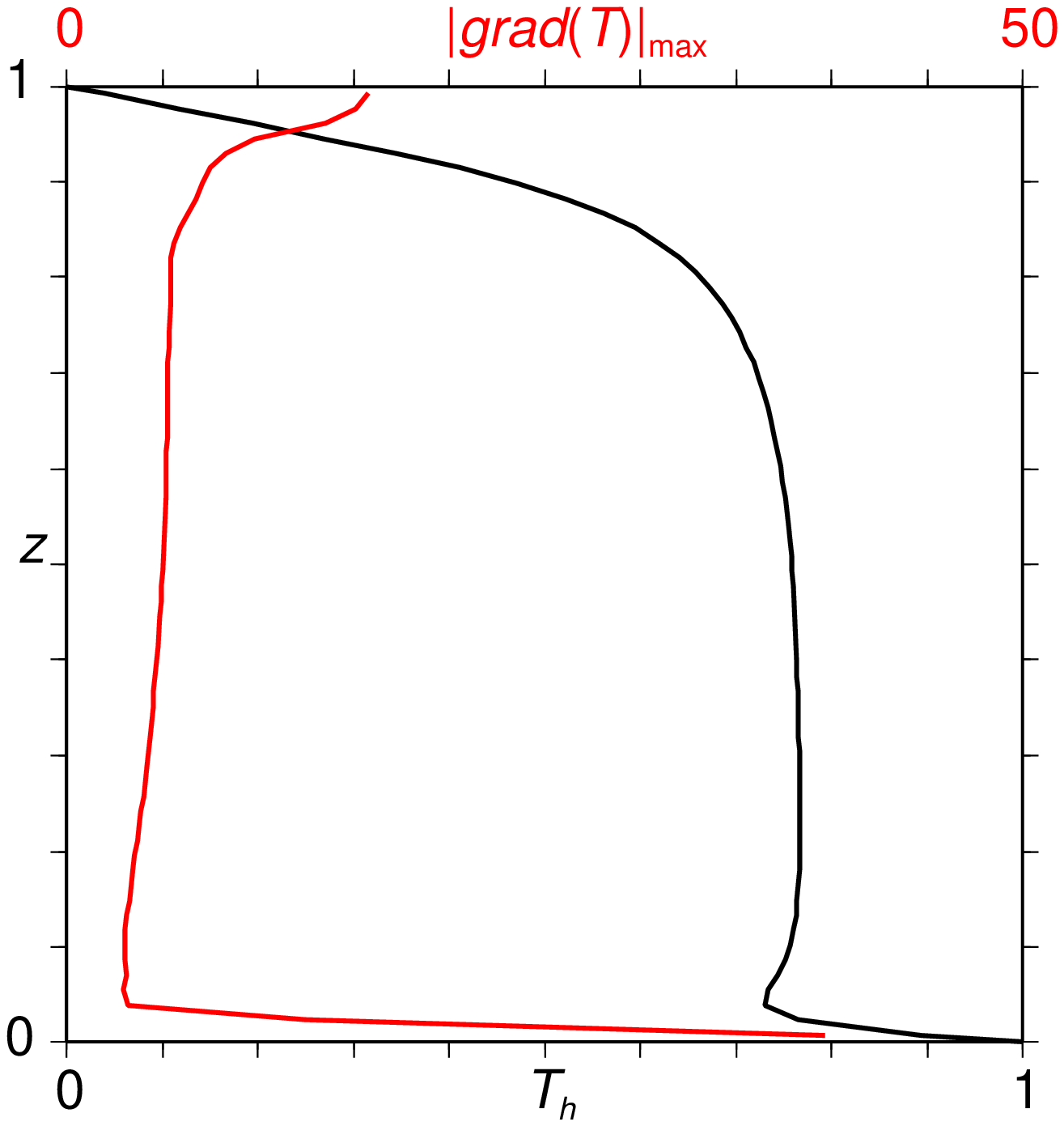}
  \\
  %\hline
  \multicolumn{2}{c}{Temperature C ($t=3.6\times10^{-2}$)} \\
  \includegraphics[height=\thisheight,keepaspectratio,clip]{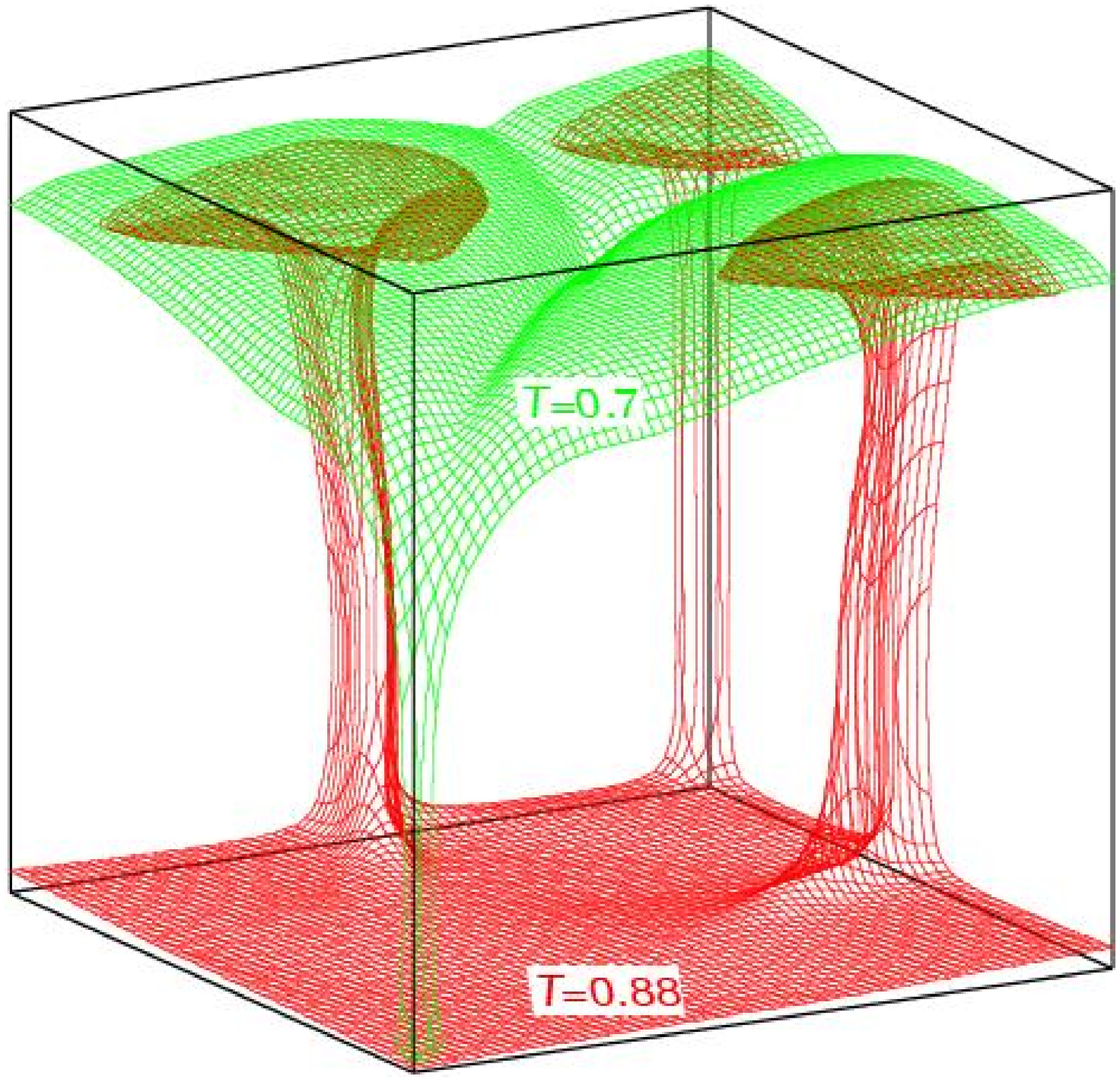}
  &
  \includegraphics[height=\thisheight,keepaspectratio,clip]{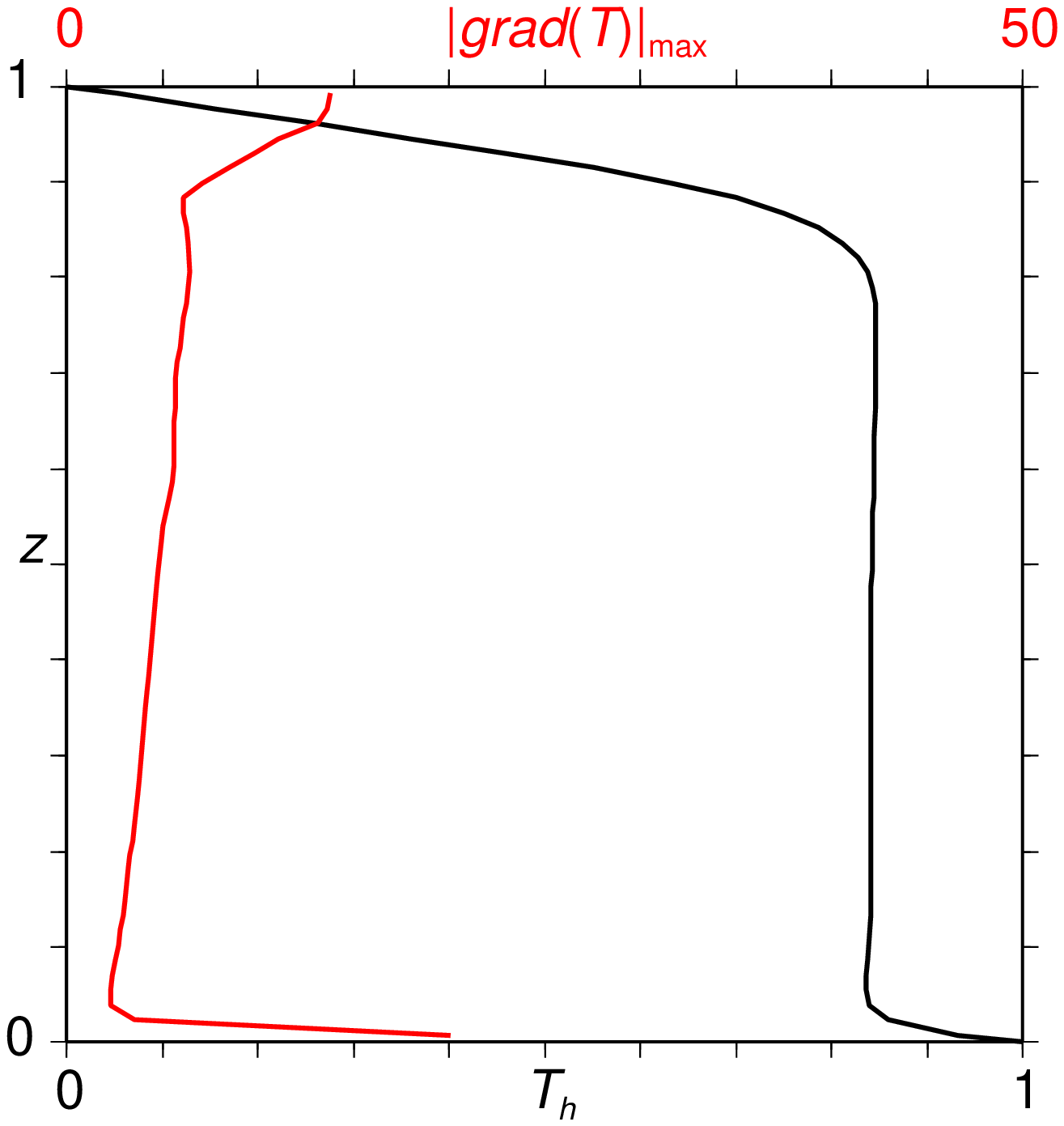}
  \\
  %\hline
 \end{tabular}
\end{center}

\clearpage
\begin{center}
 Figure \ref{conv-graph}

 \vspace{\baselineskip}
 \setlength{\thiswidth}{0.5\textwidth}
 \begin{tabular}{cc}
  (a) number of V-cycle $\nvcycle $& (b) cost of smoothing procedures $\wsmooth$\\
  \multicolumn{2}{c}{Temperature A} \\
  \includegraphics[width=\thiswidth,keepaspectratio,clip]{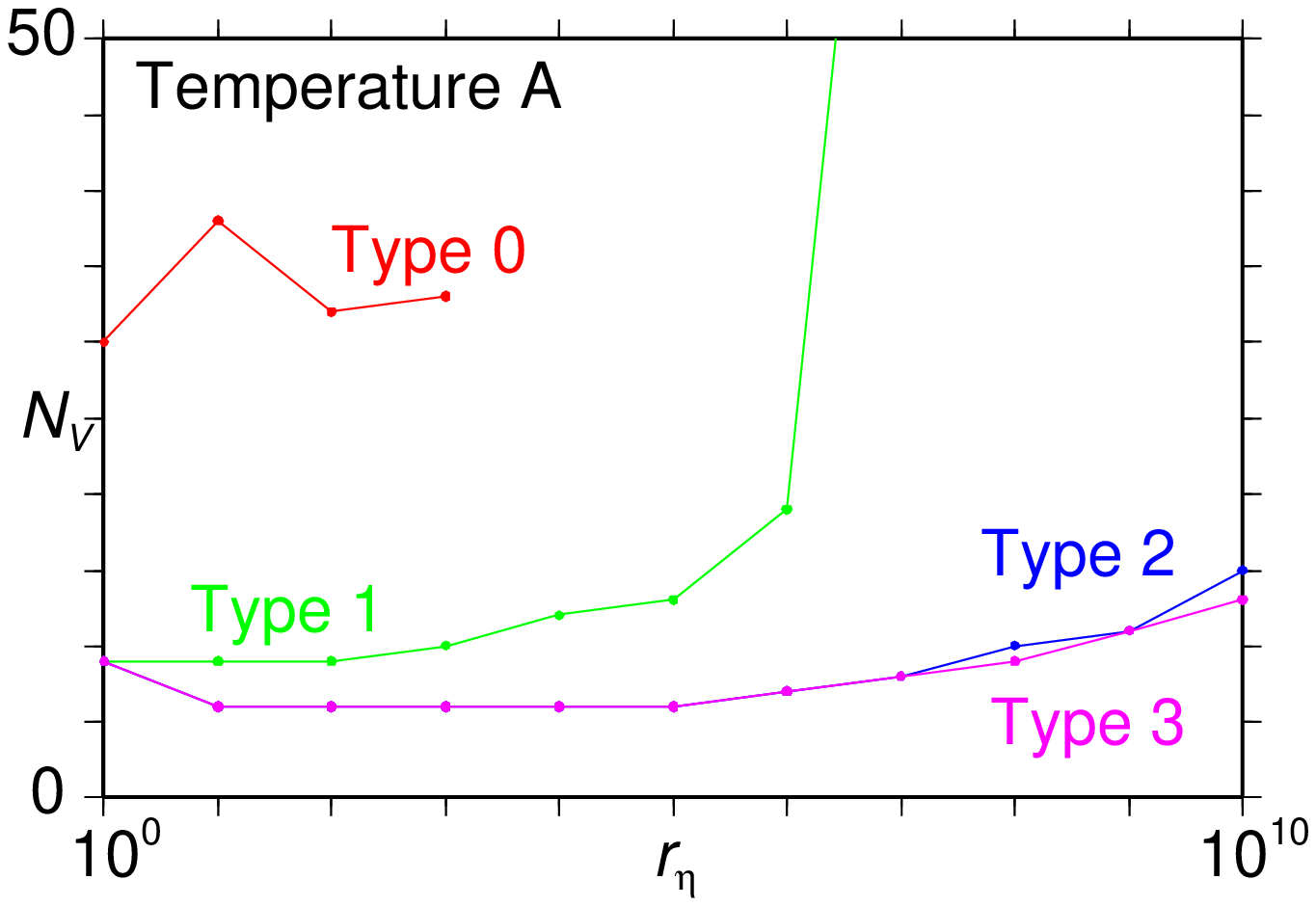}
  &
  \includegraphics[width=\thiswidth,keepaspectratio,clip]{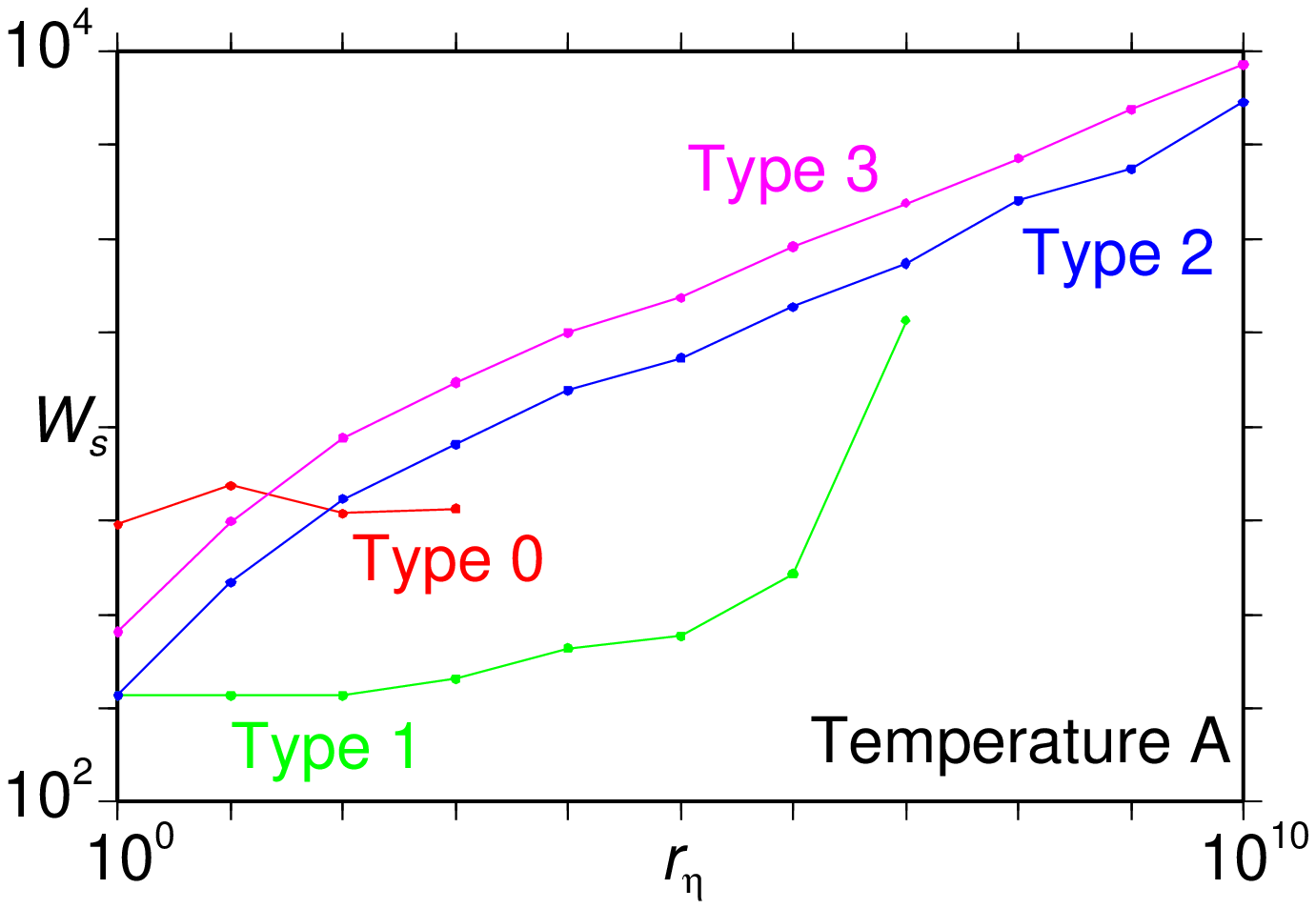}
  \\
  \multicolumn{2}{c}{Temperature B} \\
  \includegraphics[width=\thiswidth,keepaspectratio,clip]{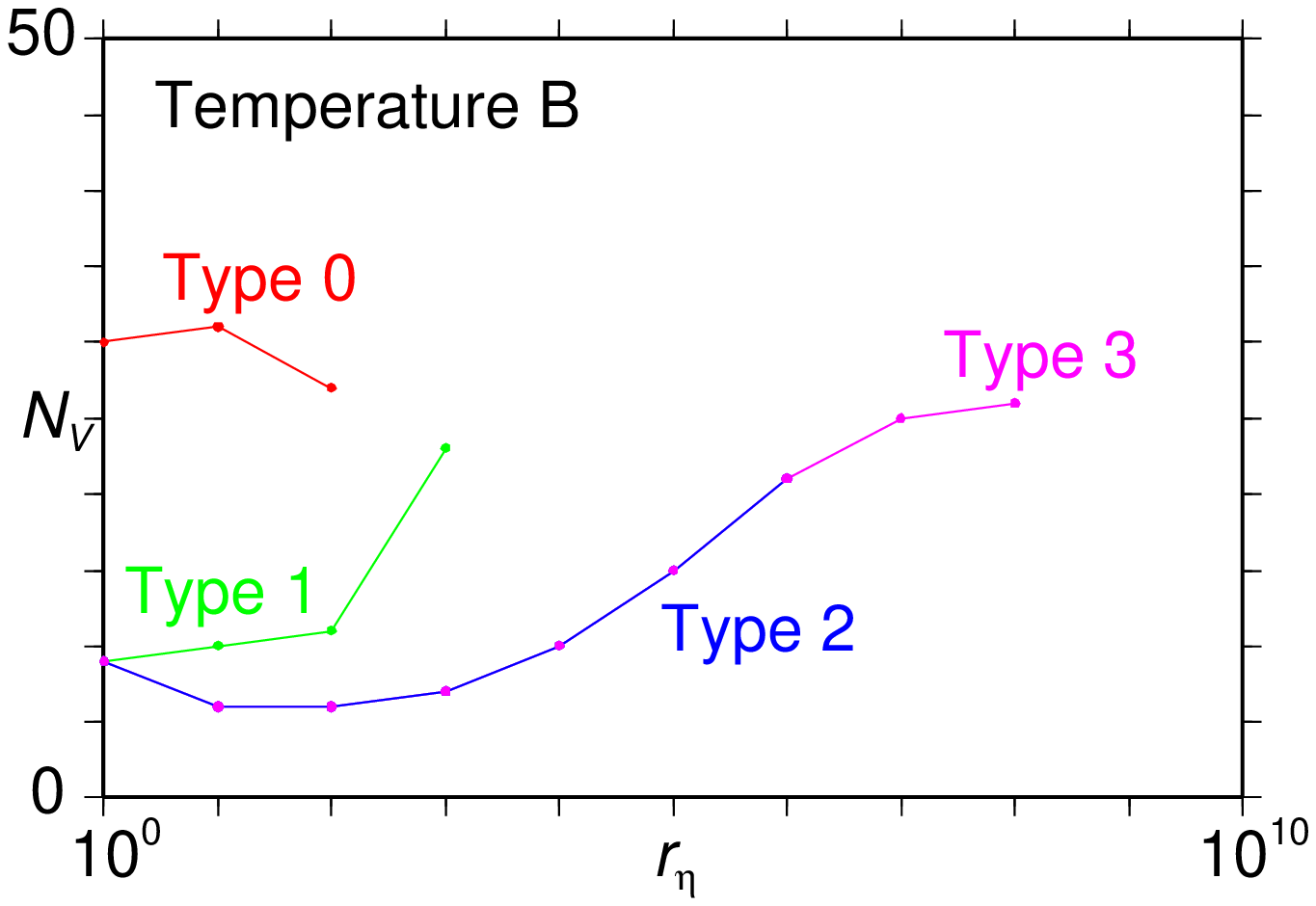}
  &
  \includegraphics[width=\thiswidth,keepaspectratio,clip]{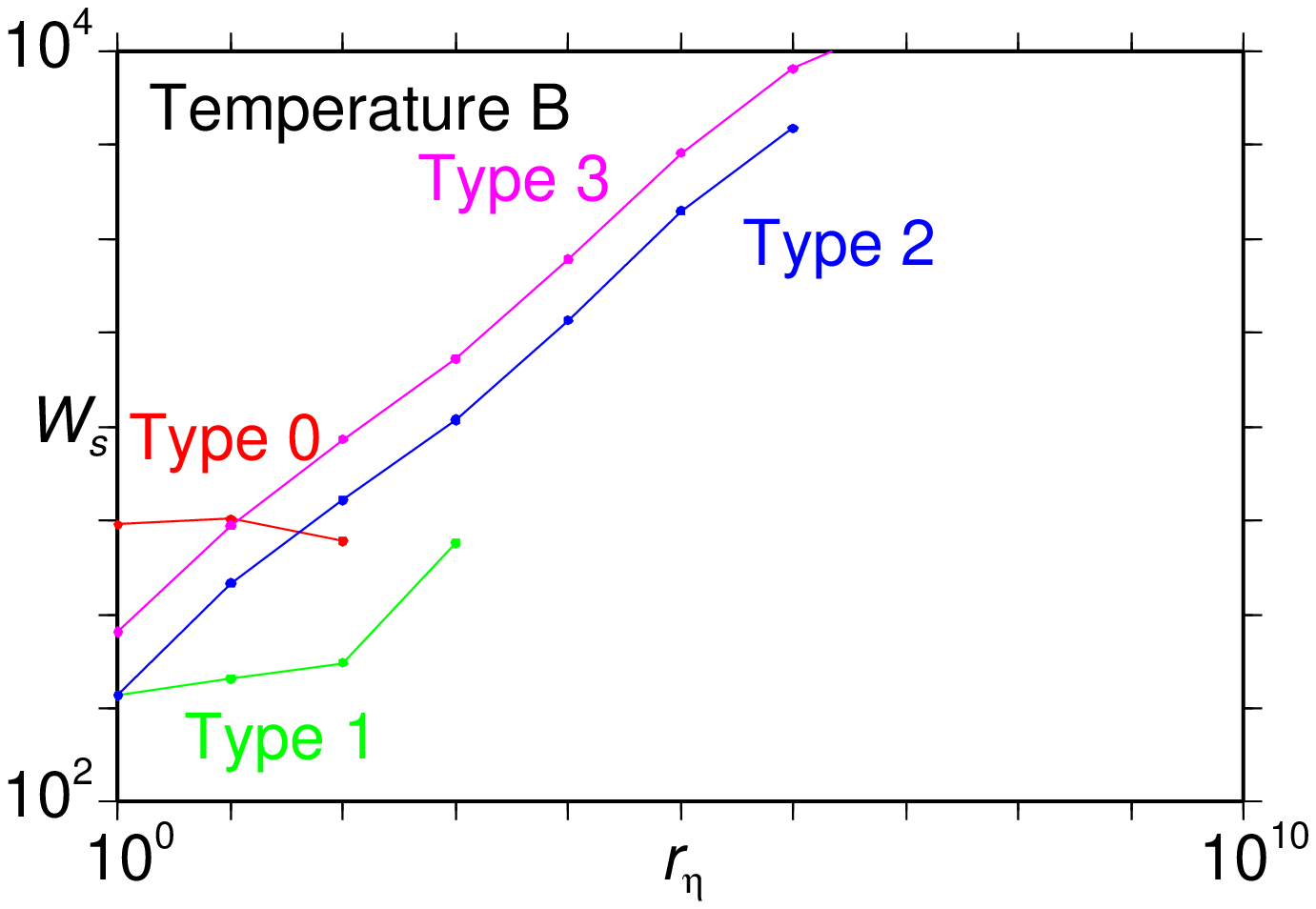}
  \\
  \multicolumn{2}{c}{Temperature C} \\
  \includegraphics[width=\thiswidth,keepaspectratio,clip]{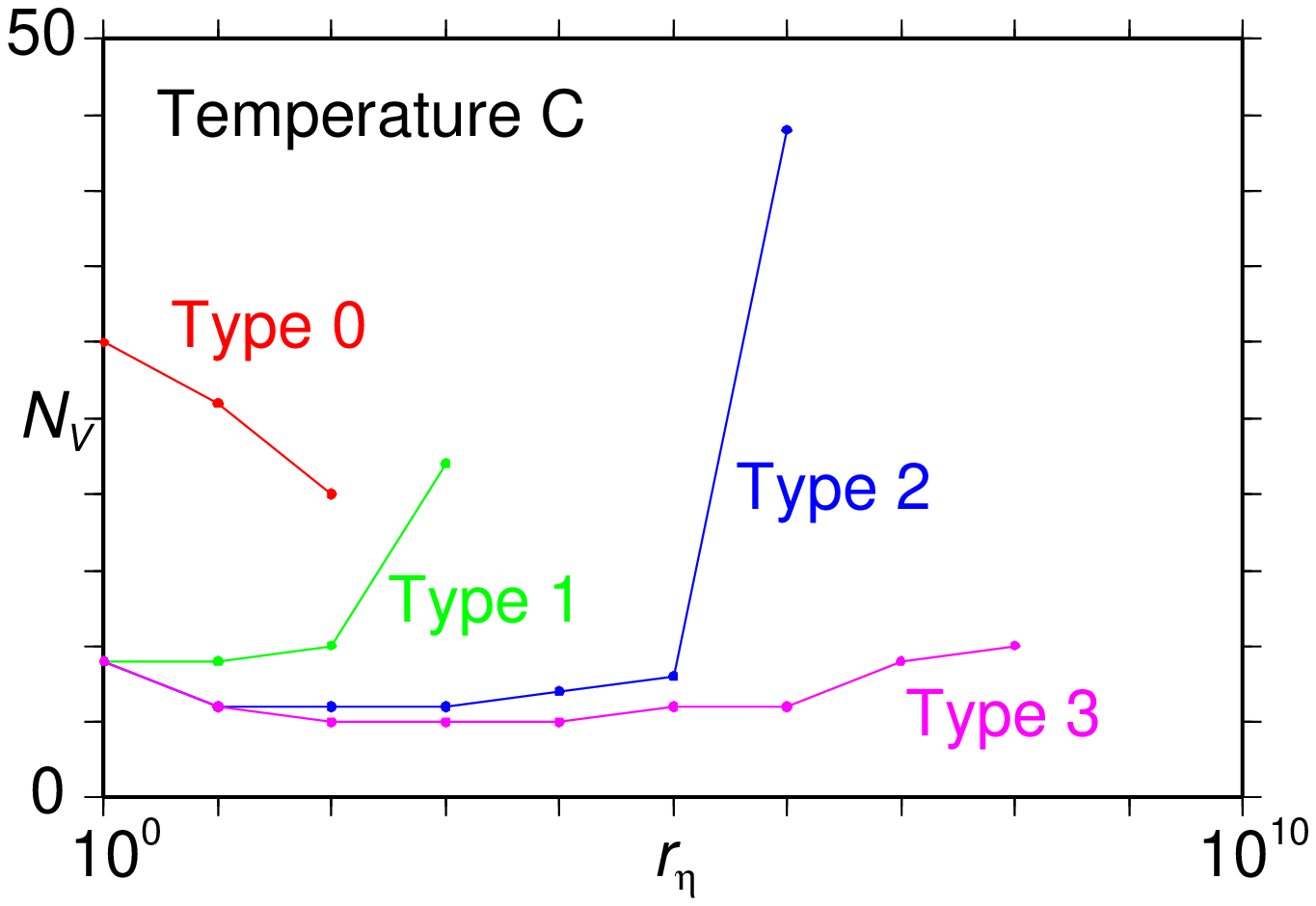}
  &
  \includegraphics[width=\thiswidth,keepaspectratio,clip]{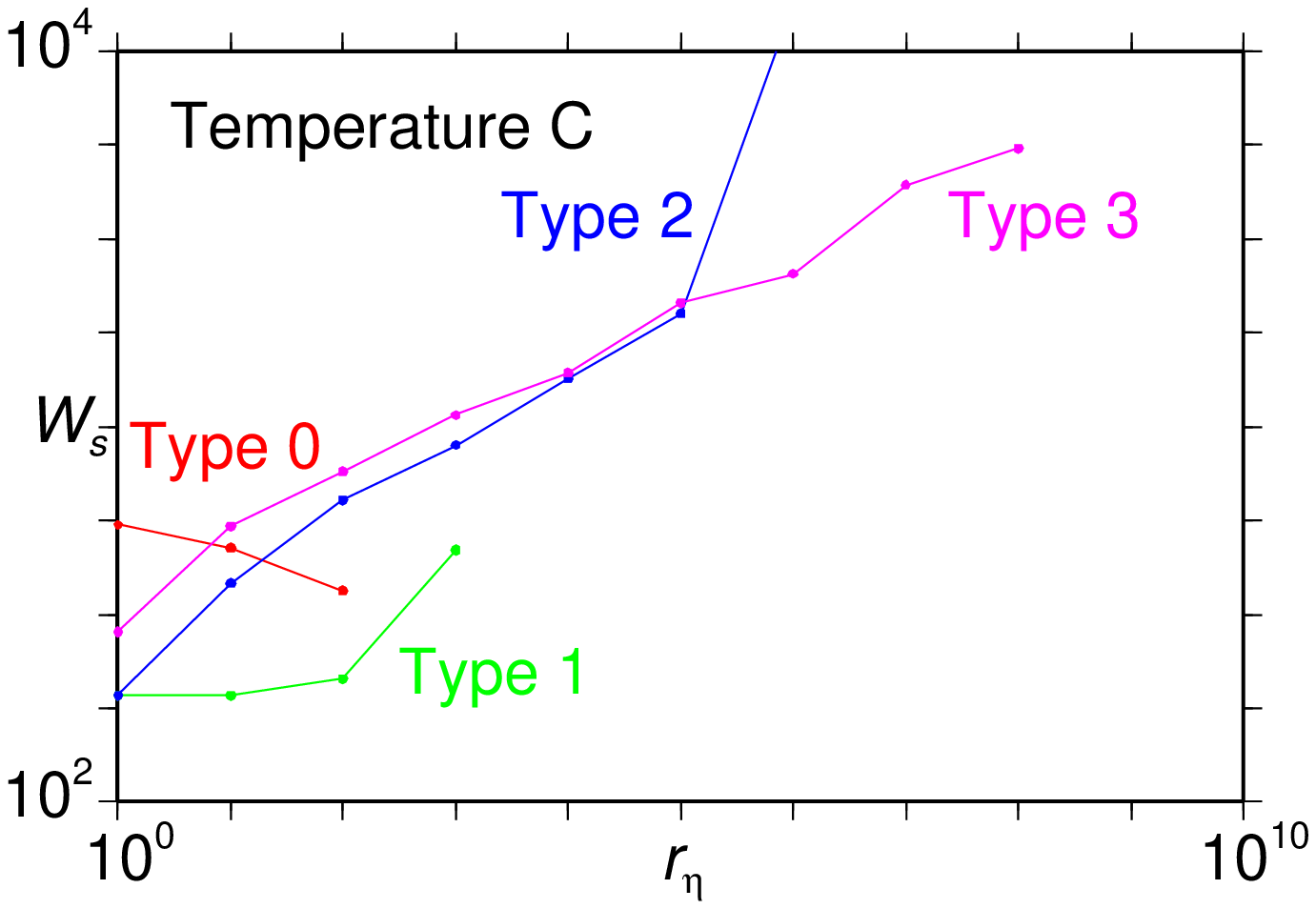}
  \\
 \end{tabular}
\end{center}

\clearpage
\begin{center}
 Figure \ref{conv-or-not}

 \vspace{\baselineskip}
 \setlength{\thiswidth}{0.5\textwidth}
 \begin{tabular}{cc}
  (a) Type 1, Temperature A, $\reta=10^4$ &
  (b) Type 0, Temperature A, $\reta=10^4$ \\
  \includegraphics[width=\thiswidth,keepaspectratio,clip]{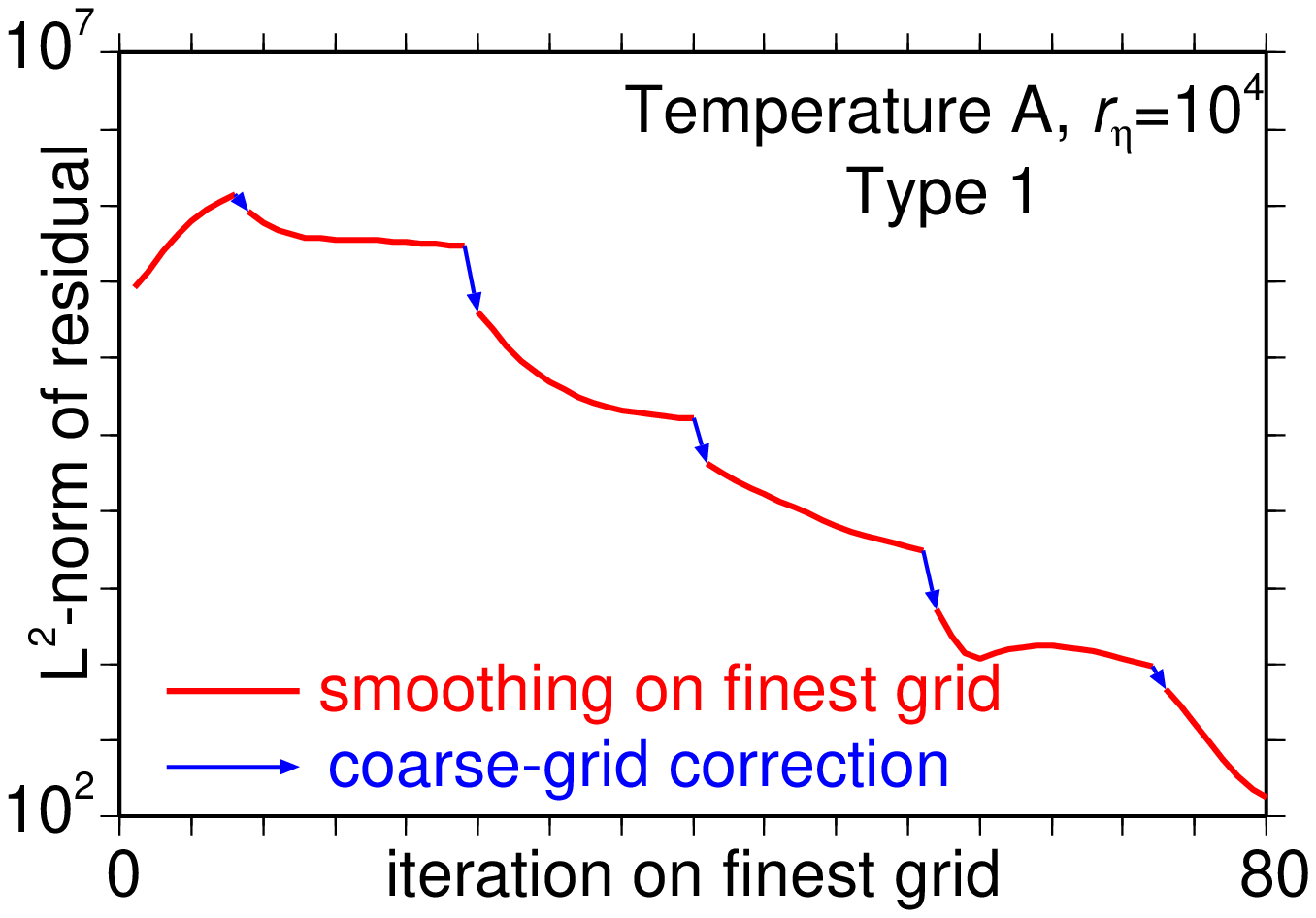}
  &
  \includegraphics[width=\thiswidth,keepaspectratio,clip]{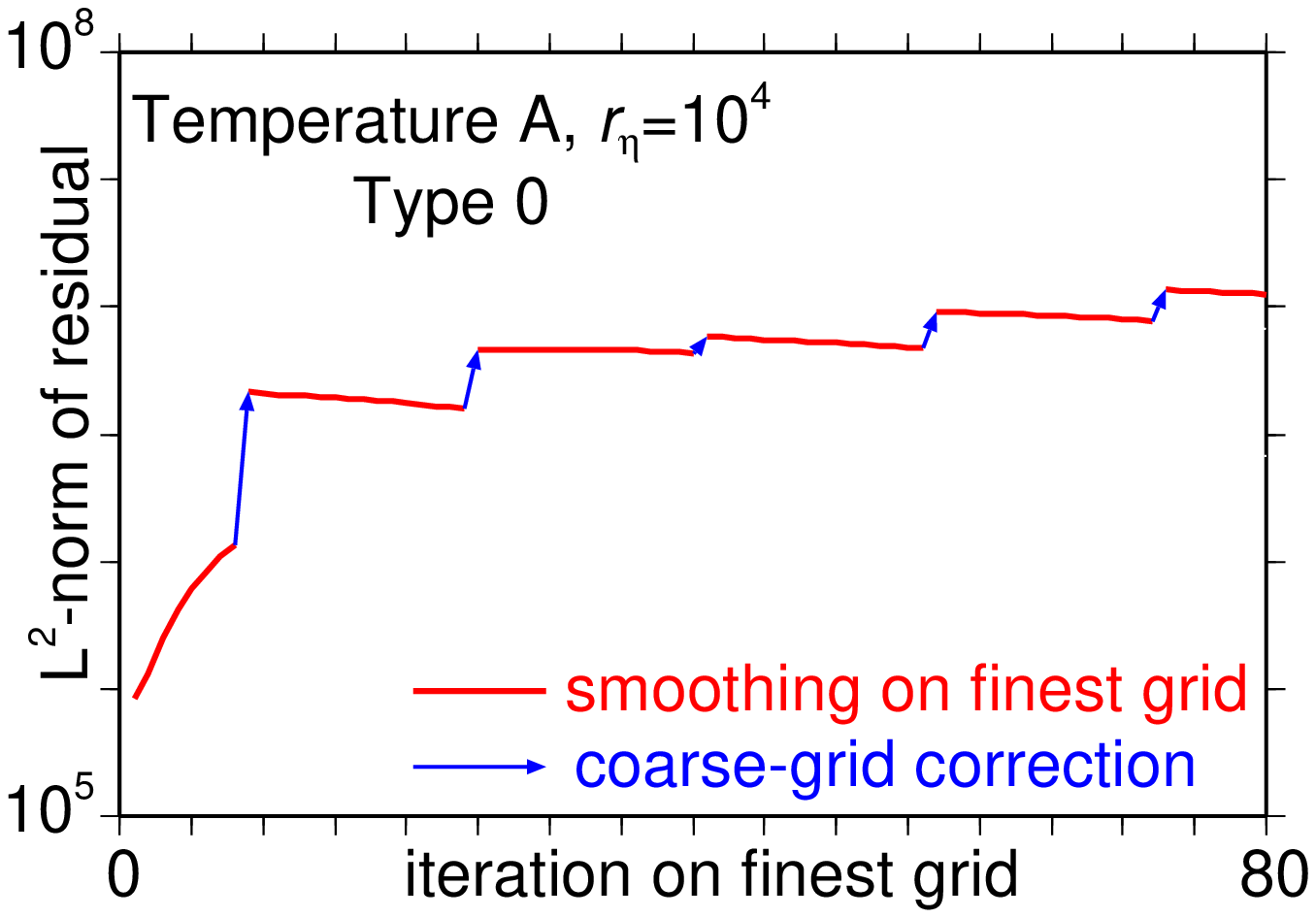}
  \\
 \end{tabular}
\end{center}
\end{document}